\documentclass[draft]{agujournal2019}
\usepackage{url} 
\usepackage{lineno}
\usepackage[inline]{trackchanges} 
\usepackage{soul}

\usepackage{comment}
\usepackage{amsmath,amssymb}

\draftfalse

\journalname{JGR - Solid Earth}

\begin{document}

\title{Localization of fast and slow slip in fault gouge and fracture energy scaling}

\authors{Dmitry I. Garagash\affil{1}, Alice-Agnes Gabriel\affil{2,3}}
\affiliation{1}{Department of Civil and Resource Engineering, Dalhousie University, Halifax, Canada}
\affiliation{2}{Scripps Institution of Oceanography, University of California San Diego, La Jolla, USA}
\affiliation{3}{Department of Earth and Environmental Sciences, Ludwig-Maximilians-Universität München, Munich, Germany}

\correspondingauthor{Dmitry I. Garagash}{garagash@dal.ca}



\begin{keypoints}
\item  Transient evolution of gouge frictional state drives localization, which in turn activates efficient thermal weakening. 
\item Fracture energy partitions into a gouge-thickness-dependent localization term and a slip-dependent post-localization term. 
\item The observed scaling of gouge thickness with natural fault size implies fault-size-dependent fracture energy scaling. 
\end{keypoints}

%
%

%
%


\begin{abstract} 

The localization of slow and fast slip in fault gouges may play a crucial role in understanding the mechanics of earthquakes and slow slip events. Here, we investigate the fracture energy accompanying this localization and the subsequent thermal weakening. We develop an analytical framework, complemented by numerical simulations, for a gouge governed by rate-and-state-dependent friction with flash-heating at high strain rate and thermal pressurization of pore fluids. The model captures the transition from initially distributed shearing to a co-seismic principal slip ``surface'' at slip $\delta_{\mathrm{loc}} \approx \gamma_c h$, and yields a decomposition of the fracture energy, $G = G_\mathrm{loc}(h) + \Delta G(\delta)$. The minimum, localization-related component $G_\mathrm{loc}$ scales with gouge thickness $h$, which in turn scales linearly with fault size.
Flash heating is activated only upon localization for fast earthquake slip, producing an abrupt strength drop, and contributing to the magnitude of $G_\mathrm{loc}$. The post-localization term $\Delta G$ increases with co-seismic slip due to efficient thermal pressurization and is insensitive to $h$. Localization is predicted to occur for both rate-weakening and rate-strengthening gouges because transient state evolution drives apparent weakening after a slip-rate increase. These results unify field, laboratory, and seismological observations of shear band thickness, critical slip, and fracture-energy scaling, and they clarify why small events can be governed by scale-dependent $G_\mathrm{loc}$ whereas large ruptures become increasingly fault-invariant as $\Delta G$ dominates. Our framework provides testable predictions for the relation of gouge thickness to lower bounds of co-seismic fracture energy, and the mechanics of slow-slip transients and fast earthquakes.

\end{abstract}

\section*{Plain Language Summary} 

Earthquakes and slow-slip events release the energy built up in the crust by tectonics to overcome fault friction. Part of that work becomes seismic waves or dissipated as heat. The rest, called fracture energy, controls how ruptures start, grow, and stop. We show that when a fault’s crushed rock (``gouge'') is sheared, deformation first spreads across the whole layer, then abruptly narrows to a thin band just a few grain diameters thick. That grain‑scale localization sets a minimum energy cost that grows with the gouge thickness, which in turn increases with fault maturity and size. Localization also turns on two heat‑driven weakening processes, flash heating at microscopic grain contacts and thermal pressurization of pore fluids, that continue to reduce strength as slip accumulates. Their slip‑dependent contribution makes the total fracture energy less sensitive to fault size for larger events. This two‑part energy budget links field, lab, and seismological observations and helps explain why small earthquakes scale differently from large earthquakes globally.

\newpage

\section{Introduction}

Localization of shear deformation manifests in nature across a wide range of spatial and temporal scales. Examples include the formation and growth of faults over geological timescale to accommodate long-term shear deformation of the Earth's crust \cite<e.g.,>{VermilyeScholz1998FaultGrowth, Scholz19, Milliner2025Localization}, the localization of discrete seismicity that is initially distributed over subsidiary-fault networks  hundreds of kilometers wide during preparation for a large earthquake \cite{Bouchon2011,EllsworthBulut2018,kato2021precursorylocalization,BenZionZaliapin2020,AlmakariBhat2026Twin}, laboratory compression of initially intact rock specimens \cite<e.g.,>{lockner1991faulting,ThompsonYoungLockner2009faultingCT,MokniDesrues1999}, 
 gouge friction experiments \cite<e.g.,>{MaroneKilgore93,mizoguchi2009,RechesLockner2010powderlubrication,KitajimaChester10,Proctor14}, 
and coseismic localization of deformation within fault cores \cite<e.g.>{Chester98, chester2005fracture, heermance2003faultzone, depaola2008localization, Chester04, savage2024localization} (Fig.~1a,b). 

Here we focus on coseismic localization on natural faults and view localization within the broader structural context of the Principal Slip Zone (PSZ), a weak natural fault gouge unit embedded within, or constituting, the fault core (Fig.~1a). 
Across faults of different cumulative displacement and maturity, the PSZ can span thicknesses from millimeters to about a meter (Fig.~1c, \cite{Chester98,WibberleyShimamoto03,Mizoguchi08, MitchellFaulkner09, zoback2011scientific}).
Field and microstructural observations further indicate that coseismic slip is accommodated on a much narrower Principal Slip Surface (PSS) embedded within a wider PSZ \cite{Chester98, chester2005fracture, Smith2011, Collettini2014} (Fig. 1b), although the extent to which exhumed fault rocks preserve a unique record of seismic slip remains an active question \cite{RoweGriffith2015}. 

Throughout experimental studies and theoretical modeling, localization is effectively synonymous to gouge weakening during slip.
However, the physical origin of the weakening that drives localization remains debated. 
 In the localization theory of Rudnicki and Rice \citeA{RuRi75} for pressure-sensitive, non-associative-plastic geomaterials (see also \cite{VardoulakisSulem95_book}), incipient localization is linked to the impending transition in plastic response from strain hardening to strain weakening. 
By contrast, several recent theoretical studies  \cite{Rice14, Platt14, platt2015TDlocalization, barrasbrantut2025localization, WangHeimisson2026PoroelasticLocalization} have attributed localization in fault gouge to shear-heating-driven thermal pressurization of pore fluids under conditions of distributed gouge shear. 
Although the latter is a plausible mechanism for fast co-seismic slip, shear heating is largely inactive in low-velocity friction experiments \cite{MaroneKilgore93, Beeler96} and in quasi-static triaxial and biaxial compression experiments on intact rock and sand, where localization is nevertheless prominent \cite<e.g.>{lockner1991faulting,ThompsonYoungLockner2009faultingCT,HanVardoulakis91,MokniDesrues1999}. This suggests that a more general weakening mechanism, or a set of mechanisms, governs incipient localization in fault gouge.

Evolution of the state of frictional contacts \cite{Dieterich79a,Ruina83} in fault gouge is likely one such ubiquitous weakening mechanism.   In response to a shear-strain-rate transient, for example imposed by a rupture front propagating along the fault, it produces apparent strain-weakening of the gouge and thereby leading to incipient localization, irrespective of whether slip is slow or fast. 
This is similar to the localization driven by hardening-to-weakening transition in strain-dependent, pressure-sensitive plasticity \cite{RuRi75, VardoulakisSulem95_book}. 
%
 However, such strain weakening by itself does not determine the internal structure of the shear band: in a purely local formulation it leads either to unbounded localization onto a surface of vanishing thickness \cite{RuRi75,VardoulakisSulem95_book} or, in patterned solutions, to a finite thickness linked to the characteristic band spacing \cite{garagash2006LocalizationPatterns}.
Previous work relating shear-band width to grain size within the framework of second-order-gradient, strain-dependent plasticity \cite<e.g.>{MuhlhausVardoulakis1987LocalizationCosserat, Vardoulakis1991gradient, VardoulakisSulem95_book, rattez2018localization_theory,rattez2018localization_num}, 
as well as within non-local granular physics \cite<e.g.>{daub2008_STZ_localization,daub2010_STZ_Rupture,pranger2022RS}, 
has provided workable theoretical frameworks for shear localization rooted in the intrinsic granularity of gouge and not requiring  thermally activated weakening at fast slip rates as the localization driver.

Once localization occurs, the resulting high shear strain rates within a narrow shear band promote efficient shear heating and activate a host of additional thermal weakening mechanisms \cite{Rice06}. These include flash heating at grain or asperity contacts 
\cite<e.g.,>{Rice06, goldsby2011flash, RempelWeaver08, BrantutPlatt2017FH}, 
which produces rapid frictional weakening over small slip distances set by the thickness of the localized shear band; thermal pressurization (TP) of pore fluids \cite<e.g.,>{Lach80, MaSm87, Rice06, badt2020TP};
thermal decomposition \cite{han2007TD, Hirose07TD,Brantut08TD,sulem2009thermaldecomposition,Brantut2010}; and melting \cite{HiroseShimamoto05melting,DiToro11}. 
These processes promote continued weakening over larger slip distances controlled by the hydro–thermo–chemo-poroelastic properties of the gouge and are largely independent of PSZ gouge thickness.

This co-event shear localization may later be reversed. Shear strain rate may delocalize within the gouge during the inter-event period through gouge healing and/or creep \cite{zoback2011scientific}, or during the late stages of slip through through deceleration, pore-fluid and heat diffusion, or rheological changes such as melting. Laboratory slip-hold-slip friction experiments  \cite{beeler2022DcAgeDependent, Bedford2023} and observations from natural earthquake gouge \cite{savage2024localization} provide support for both inter-event and late-stage co-event delocalization, respectively. This suggests that natural slip transients may commonly initiate from a relatively delocalized state, with coseismic re‑localization following the onset of dynamic slip.
This framework is relevant across the full spectrum of observed fault slip behavior, from slow slip events (SSEs) to fast earthquakes \cite{RogersDragert2003, Ide2007, PengGomberg2010, BehrBurgmann2021}.

Fracture energy is a key energetic characteristic of fault weakening during slip and governs rupture initiation, propagation, and arrest for a given fault and surrounding bulk rock loading \cite<e.g.>{PalmerRice73, AbercrombieRice05, weng2019dynamics, kammer2024earthquake, Cocco2023}. Fracture energy in both natural and laboratory earthquakes is known to scale with ruptured fault size and coseismic slip \cite<e.g.>{AbercrombieRice05, ViescaGaragash15, Nielsen2016G,gabriel2024_FractureEnergy}. Because shear localization is synonymous with weakening, and therefore likely governs dynamic frictional weakening and the fracture energy dissipated during coseismic slip, the goal of this study is to interpret fracture energy observations in the context of the physics of localization and fault gouge properties, in particular their scaling with fault maturity and size.

Building on seismological inferences of coseismic fracture energy, \citeA{gabriel2024_FractureEnergy} showed that the total fracture energy can be decomposed into two additive parts,
\begin{equation}
G \;=\; G_c (R)\;+\; \Delta G(\delta),
\end{equation}
where $G_c(R)$ is a minimum, ``\emph{small‑slip}" contribution that we will associate here with the mechanical work of coseismic localization and that scales with a fault size $R$, here linked to gouge/PSZ thickness $h(R)$ (Fig. 1c). $\Delta G(\delta)$ is a ``\emph{large‑slip}", continuing‑weakening contribution that increases with coseismic slip $\delta$, for example due to TP acting on an already localized slip surface or PSS. This framework predicts a transition from fault‑size–dependent behavior at small slip, dominated by $G_c$, to fault‑invariant but slip‑dependent behavior at larger slip, dominated by $\Delta G$, thereby providing a physical basis for differences between small and large earthquake ruptures.

In the following, we analyze strain localization and subsequent weakening in fluid‑saturated fault gouge across a wide range of PSZ thicknesses $h$ (millimeters to meters), representative of faults ranging from immature fault systems to highly mature plate‑boundary faults and spanning a wide range of sizes $R$ (Fig.~1c). Using an analytical and numerical framework that incorporates full rate‑and‑state friction together with thermal weakening, we quantify how localization governs the partitioning of fracture energy into $G_c(R)$ and $\Delta G(\delta)$. 


\section{Model}

This section presents a model and the underlying physics of transient shear of a gouge layer and localization in response to an imposed sliding velocity $V(t)$ history, as illustrated in Fig. \ref{fig:sketch}. 

\subsection{Stress distribution in fault gouge}

Neglecting inertia of distributed slip within a gouge layer on the scale of the gouge thickness $h$ \cite{Rice14}, results in a spatially uniform normal stress ($\sigma$) and shear stress ($\tau$) distribution within the gouge, $|y|<h/2$, as (Fig. \ref{fig:sketch}a):
\begin{equation}
\sigma(y,t)=\sigma,\quad\tau(y,t)=\tau(t)\,.\label{eq}
\end{equation}
We neglect dilation of the gouge resulting in time-invariant normal stress in Eq. (\ref{eq}).

\subsection{Pore pressure in fluid-saturated, shear-heated fault gouge}
Pore pressure $p(y,t)$ in fluid-saturated gouge can evolve in time and across the gouge layer in response to the thermal pressurization driven by the gouge shear-heating and the related exchange of the fluid and heat between gouge and surrounding fault rocks \cite<e.g.,>[]{Lach80,Rice06}. To describe the coupled thermal and hydraulic response of the fault gouge, we formulate the governing equations for temperature and pore pressure evolution in fluid-saturated, shear-heated fault gouge, following \cite{Rice06,RempelRice06,Garagash12}: 
\begin{equation}
\frac{\partial T}{\partial t}=\frac{1}{\rho c}\tau\dot{\gamma}+\alpha_{\mathrm{th}}\frac{\partial^{2}T}{\partial^{2}y}\,,\qquad 
\frac{\partial p}{\partial t}=\Lambda\frac{\partial T}{\partial t}+\alpha_{\mathrm{hy}}\frac{\partial^{2}p}{\partial^{2}y}\,,
\end{equation}
Here $\rho c$ is the specific heat capacity,  $\alpha_{\mathrm{th}}$ and $\alpha_{\mathrm{hy}}$ are thermal and hydraulic diffusivity, respectively. $\Lambda=(\partial p/\partial T)_{\mathrm{undr}}$ is the undrained thermal pressurization coefficient, quantifying the increase of pore pressure per unit increment of temperature rise due to the differential thermal expansion between the pore fluid and the pore space in the gouge and surrounding rock. Shear heating acts as a heat source $\tau\dot{\gamma}$  within the sheared gouge $|y|<h/2$, and is zero outside,  i.e., $\dot{\gamma}=0$ for $|y|>h/2$.

We assume more efficient diffusion of fluids compared to heat $\alpha_{\mathrm{hy}}\gg\alpha_{\mathrm{th}}$ \cite{Rice06, Garagash12}. This allows us to neglect heat conduction, and approximate the evolution of pore pressure $p$, or equivalently, of the effective stress $\bar{\sigma}=\sigma_{o}-p$, under nearly adiabatic conditions, as:
\begin{equation}
-\frac{\partial\bar{\sigma}}{\partial t}=\frac{\Lambda}{\rho c}
\tau
\dot{\gamma}-\alpha_{\mathrm{hy}}\frac{\partial^{2}\bar{\sigma}}{\partial^{2}y} \,.
\label{dp}
\end{equation}

\subsection{Frictional rheology of gouge}\label{sec:fric}

To characterize the frictional strength evolution within fault gouge and its implications for strain localization and dynamic weakening, we adopt a rate-and-state-dependent rheological framework incorporating both local and non-local shear deformation effects.
\subsubsection{Rate-and-State-Dependent Gouge Friction}
Here, we introduce the basic equations, rate-and-state friction rheology, and definitions of strain localization scales.
Within the sheared gouge, the stress-ratio $\tau/\bar{\sigma}$ is equal to the gouge friction
\begin{equation}
\tau/\bar{\sigma}=f\,,\label{tau}
\end{equation}
where $\bar{\sigma}=\sigma-p$ is the effective stress normal to the fault and $p$ is the pore-fluid pressure.  We note that stresses ($\tau$ and $\sigma$) are spatially uniform across the gouge due to mechanical equilibrium. Therefore, any spatial non-uniformity of gouge friction can only arise from variations in pore fluid pressure induced by thermal pressurization and diffusion.

Friction evolution follows a rate-and-state-dependent rheology \cite{Rice83}, adopted for distributed shear deformation:
\begin{equation}
f=f_{\mathrm{ref}}+a\ln\dot{\gamma}/\dot{\gamma}_{\mathrm{ref}}+\Theta\label{f}\,,
\end{equation}
where $\dot{\gamma}=\partial\gamma/\partial t$ is the local shear strain rate and $\Theta$ is the state variable. 
The evolution of $\Theta$ follows the `slip (strain) law' \cite{Ruina83}:
\begin{equation}
\frac{\partial\Theta}{\partial t}=-\frac{\dot{\gamma}}{\gamma_{c}}\,\left(f-f_\mathrm{ss}(\dot{\gamma})\right)\,,\label{Theta}
\end{equation}
where $\gamma_{c}$ is the characteristic ``state-evolution'' strain required for the local frictional state to approach steady-state during distributed gouge shearing \cite{sleep1997RSfrictionLocalization,MaroneKilgore93,marone2009Localization,FerdowsiRubin2020granulargouge} and $f_\mathrm{ss}(\dot{\gamma})$ is the steady-state friction at strain rate $\dot{\gamma}$. 

The value of $\gamma_c\approx 0.13$ has previously emerged from discrete element simulations of gouge shear \cite{FerdowsiRubin2020granulargouge}. 
Alternatively, $\gamma_c$ can be estimated from combining existing experimental observations of localization strain $\gamma_\mathrm{loc}$ in granular materials with the solution of our model which links $\gamma_\mathrm{loc}$ to $\gamma_c$.
In gouge direct shear experiments \cite{MaroneKilgore93,Marone98} and sand biaxial compression experiments \cite{HanVardoulakis91,MokniDesrues1999}, localization strains of $\gamma_\mathrm{loc}\sim0.05\,\text{-}\,0.1$ were inferred.
On the other hand, as we show in Sec. \ref{sec:results} and in Fig. \ref{fig:collapse}a,c, our model suggests that $\gamma_\mathrm{loc}$ scales with $\gamma_c$, as $\gamma_\mathrm{loc}\sim (0.1\,\text{-}\,1)\times\gamma_c$.  
Taken together, this suggests $\gamma_c$ falls in the range of $\sim$ 0.1 to 1.


The strain rate is bounded by a minimum value corresponding to the fully de-localized slip distributed across the entire gouge layer thickness, $\dot{\gamma}\approx V/h$, and a maximum value corresponding to fully localized slip within a thin gouge sub-layer of a much smaller thickness $h^*$, with $h^*\ll h$, that is expected to scale with the grain size \cite{Rice06}, $\dot{\gamma}\approx V/h^*$. In both limiting cases, Eq. (\ref{Theta}) reduces to the classical slip-law \cite{Ruina83}, $\partial\Theta/\partial t=-(V/d_c)\,(f-f_\mathrm{ss}(V))$, but with drastically different state-evolution slip distances $d_c$, given respectively by $\gamma_c h$ and $\gamma_c h^*$. 
Here, $h^*$ represents the minimum possible, irreducible localization band thickness linked to gouge granularity; however, the actual localization thickness can be larger due to localization-limiting processes beyond grain-size constraints.  
For instance, in studies by  \cite{Rice14,Platt14}  localization was limited by the assumed frictional velocity-strengthening with a rheology only depending on slip rate and not on the state variable, and gouge granularity was not explicitly incorporated or discussed. 

\subsubsection{Steady-state friction and flash heating}
Differentiating Eq. (\ref{f}) with respect to time and substituting Eq. (\ref{Theta}), we recover an expression for the friction-time-rate $\partial f/\partial t$,
\begin{equation}
\frac{\partial f}{\partial t}=a\frac{\partial\ln\dot{\gamma}}{\partial t}-\frac{\dot{\gamma}}{\gamma_{c}}\left(f-f_\mathrm{ss}(\dot{\gamma})\right)\label{df}
\end{equation}
We extend the definition of steady-state friction from prior studies \cite{Rice06,Beeler08,Noda09,dunham2011planar} as follows
\begin{equation}
f_\mathrm{ss}(\dot{\gamma})=\chi(\dot{\gamma})\,f_\mathrm{LV}(\dot{\gamma})+(1-\chi(\dot{\gamma}))\,f_{w},\qquad\chi(0)=1,\quad\chi(\infty)=0\,.
\label{ss}
\end{equation}

The ``low-velocity'', or more appropriately ``low-shear-rate'', response for $\chi\approx1$ corresponds to the classical logarithmic rate-dependence 
\begin{equation}
f_\mathrm{LV}(\dot{\gamma})=f_{\mathrm{ref}}+(a-b)\ln\dot{\gamma}/\dot{\gamma}_{\mathrm{ref}}
\label{fLV},
\end{equation}
where the reference strain rate $\dot{\gamma}_{\mathrm{ref}}=V_{\mathrm{ref}}/h^*$ is defined in terms of the reference slip velocity $V_{\mathrm{ref}}$ for fully localized slip. Thus, (\ref{fLV}) reduces to the classical relation in terms of slip rate $V$ \cite{Dieterich79a, Marone98}, $f_\mathrm{LV}=f_{\mathrm{ref}}+(a-b)\ln V/V_{\mathrm{ref}}$, in the case of complete slip localization, $\dot{\gamma}=V/h^*$. The low-velocity rock gouge friction is $\sim$ 0.6 - 0.8 \cite<e.g.,>[]{paterson2005experimental}.
The ``high-velocity'', or rather ``high-strain-rate'', frictional response with $\chi\approx0$ corresponds to the fully-weakened, flash-heated friction value $f_{w}\lesssim 0.1$. 

The transition from low-to-high strain-rate friction behavior is governed by the function $\chi(\dot{\gamma})=\left(1+(\dot{\gamma}/\dot{\gamma}_{w})^{m}\right)^{-1/m}$, as adopted from \cite{dunham2011planar}.  The flash-heating strain-rate threshold  $\dot{\gamma}_{w}=V_{w}/h^*$ is defined in terms of the flash-heating slip velocity $V_{w}$ for the fully-localized slip conditions.
The exponent $m\ge1$ controls the smoothness of the transition as strain-rate $\dot{\gamma}$ increases past $\dot{\gamma}_{w}$, such that larger values of $m$ produce a sharper transition.

\subsubsection{Constraints on fully-localized shear band thickness $h^*$ and flash-heating activation strain-rate $\dot{\gamma}_{w}$}

Friction experiments on artificial gouges with varied grain-size distributions \cite{MaroneKilgore93} show values of state-evolution slip distances $d_c$ ranging between 1 and 10 $\mu$m when measured at sufficiently large cumulative strain presumably after undergone strain localization.
Given the relation $d_c=\gamma_c h^*$ and assuming the state-evolution strain $\gamma_c$ to be in the range between 0.1 to 1, the fully-localized shear band thickness $h^*$ can be indirectly estimated to fall in the range from 10 to 100 $\mu$m. 

This estimate is consistent with direct experimental observations showing that shear bands typically span from a few to a few tens of grains across (see review by \citeA{Rice06} references therein). Natural ultracataclastic gouges in principal slip zones (PSZs, Fig.~1a) of mature faults commonly have  a median grain size of $\sim 1~\mu$m, with distributions spanning from nanometers to tens of microns \cite{Wilson05}. Consequently, they may host very thin localized bands ($h^* \sim 10~\mu$m). Microstructural observations from the Punchbowl fault principal slip surface (PSS) are consistent with such extreme localization \cite{Chester98, chester2005fracture} (Fig. 1b). In contrast, localization bands in many artificial gouge experiments are typically thicker  ($h^* \gtrsim 100~\mu$m, \cite<e.g., >{Proctor14}) commensurate with their larger median grain size ($\gtrsim 10~\mu\mathrm{m}$). This also suggests that  $h^*$ in artificial gouge may evolve with ongoing comminution.

Flash heating is expected to occur once a critical contact-scale strain rate, $\dot\gamma_w $, is exceeded \cite{Beeler08, Platt14, Proctor14}. This implies that the corresponding critical slip velocity $V_w$ scales with the localized band thickness $h^*$ as $\dot\gamma_w = V_w/h^*$. 
Laboratory constraints suggest $V_w \sim 0.1$~m\,s$^{-1}$ on bare rock surfaces \cite{KohliGoldsby11, goldsby2011flash, DiToro11}, whereas sheared rock gouges require higher values, $V_w \sim 1$~m\,s$^{-1}$ \cite{Proctor14, YaoMaPlatt15}. For bare surfaces, $h^*$ can be approximated by the asperity size observed in microstructures ($h^* \sim 10~\mu$m), whereas artificial gouge experiments report localized bands with $h^* \sim 100~\mu$m after substantial grain-size reduction to $\lesssim 20~\mu$m. Both cases yield a consistent estimate of $\dot\gamma_w \sim 10^{4}$~s$^{-1}$, supporting the notion that the onset of flash heating is fundamentally controlled by a strain‑rate threshold. Consequently, \emph{flash-heating} is expected to remain \emph{inactive} even at seismic slip rates of $V\gtrsim 1$ m/s, as long as shear deformation remains \emph{distributed}, $\dot{\gamma}\sim V/h$, within gouge layers of thickness $h\gtrsim 1$ mm. We therefore anticipate that flash-heating-induced weakening can occur only after shear localization and is absent during delocalized slip.

In the following analysis, unless stated otherwise, we adopt natural ultracataclastic gouge and bare-rock surface constraints \cite{KohliGoldsby11, goldsby2011flash, DiToro11} and set the localized shear-band thickness to $h^*=10\,\mu$m. The critical flash-heating activation strain rate is set to $\dot\gamma_w = 10^{4}$~s$^{-1}$  and the corresponding critical slip velocity is $V_w=\dot\gamma_w h^*=0.1$ m/s.

\subsection{Grain-scale physics and structure of a localized shear band}

The following discusses the grain-scale physics that prevents the otherwise ``run-away'' (unbounded) localization onto a mathematical plane, thereby defining the finite thickness and internal structure of the localized shear band.

\subsubsection{Impending run-away localization in a local frictional rheology}

Although in the frictional rheology formulation in Sec. \ref{sec:fric} we invoke gouge granularity and the existence of a minimum, irreducible localized shear thickness $h^*$, this framework remains inherently local, meaning it depends only on the strain rate and state variables defined at a single point. Consequently, it is expected to lead to an unabated localization process inconsistent with the irreducible finite shear band thickness $h^*$.  
Under steady-state rate-weakening conditions, either due to a low-strain-rate weakening regime ($a-b<0$) or due to localization activation of flash-heating weakening at sufficiently high imposed slip velocity, this local framework would predict unbounded shear localization, collapsing onto a mathematical plane and resulting in infinite strain rates. Furthermore, even in the case of steady-state rate-strengthening gouge friction, the inherent weakening driven by the evolution of the frictional state following a transient increase in slip velocity also drives the localization onto a mathematical plane.  

\subsubsection{Non-local friction rheology}

In reality, the gouge granularity (finiteness of grain sizes) frustrates the otherwise unbounded localization. This manifests as a departure from the purely local gouge rheology at the large spatial gradients in strain rate intrinsic to the localization process. The granularity is expected to lead to a finite thickness $h^*>0$ of the localized shear band and non-local rheological effects exercised during localization \cite<e.g.,>[]{daub2008_STZ_localization,rattez2018localization_theory, rattez2018localization_num,pranger2022RS,Casas2025}, as also documented in granular physics \cite<e.g.,>[]{ErtasHalsey02,Bonamy02,Pouliquen04,Mills08}.

This non-locality originating from the fine-scale granularity of the fault gouge can be accounted for by introducing a non-local strain rate measure \cite<e.g.,>[]{pranger2022RS,henann2013,bouzid2013nonlocal}:
\begin{equation}
\dot{\gamma}_\lambda(y,t)=\left( 1+\lambda^2 \frac{\partial^{2}}{\partial^{2}y}\right) \dot{\gamma}(y,t)\,,
\label{non-loc}
\end{equation}
where $\lambda$ is the spatial scale of non-locality, intuitively associated with grain or asperity size. 
This strain-rate measure approximately averages the local strain rate over a spatial neighborhood of size $\sim \lambda$ . 

Following the approach of \citeA{pranger2022RS} for a rate-and-state gouge friction framework, non-local effects can affect both the steady-state friction value $f_\mathrm{ss}$, Eqs. (\ref{ss}-\ref{fLV}), and the evolution of gouge toward steady-state, Eq. (\ref{Theta}). Thus, a modified, non-local formulation is obtained by replacing the local strain rate $\dot\gamma$ with the non-local strain rate $\dot{\gamma}_\lambda$ in these equations.

\subsubsection{Structure and thickness of a shear band}

An approximate solution for the structure $\dot\gamma(y)$ and thickness $h^*$ of the localization band, $|y|\le h^*/2$, can be obtained under two assumptions: (i) frictional steady-state, $\tau\approx f_\mathrm{ss}(\dot\gamma_\lambda)\,\bar\sigma$; and (ii) drained conditions, i.e., the effective stress $\bar\sigma$  (pore pressure) is approximately uniform within the localization band. 

The first assumption of frictional steady-state is expected to hold as long as the sliding velocity remains approximately constant over slip distances of the order of $d_c=\gamma_c h^*\sim 1$ $\mu$m. 
The second assumption of a drained shear band applies when the hydrothermal diffusion timescale, $t^*=h^{*2}/\alpha_\mathrm{hy}$, is much smaller than the thermal pressurization timescale, $t_\mathrm{heat}=\rho c / (\Lambda \dot\gamma)\sim \rho c h^*/ (\Lambda V)$, \cite{Garagash12}. Taking $h^* \sim 10\,\mu$m, and assuming typical hydrothermal parameters of a ``damaged'' ultracataclastic gouge \cite{Rice06}, $\rho c/\Lambda \sim 10$ and $\alpha_\mathrm{hy} \sim 10^{-5}$ m$^2$/s, we estimate the two timescales to be $t^* \sim 10^{-5}$ s and $t_\mathrm{heat}\sim (100\,\mu\text{m})/V$. This suggests that assuming drained shear band conditions is valid at slip rates $V\gg 100\,\mu$m/s.

Under these assumptions, stress equilibrium dictates that the non-local strain rate $\dot\gamma_\lambda$, Eq. (\ref{non-loc}), is constant across the band. Integrating Eq. (\ref{non-loc}), assuming that the strain-rate vanishes outside the band, $\dot\gamma=0$ for $|y|>h^*/2$, and corresponding kinematic condition $\int_{-h^*/2}^{h^*/2}\dot\gamma dy=V$, we find 
\begin{equation}
    |y|\le h^*/2:\qquad \dot\gamma=\frac{V}{h^*}\left(1+\cos\frac{y}{\lambda}\right)
\quad\text{with}\quad
h^*=2\pi\lambda\,.
\label{non-loco}
\end{equation}

The corresponding constant value of the non-local strain rate is $\dot\gamma_\lambda=V/h^*$.

Above solution for the shear band structure demonstrates that the localized shear band thickness $h^*$ emerging from the non-local framework is approximately $2\pi \lambda$. 
Thus, if $\lambda$ in Eq. (\ref{non-loc}) is interpreted as the grain size, the resulting localization shear band thickness would span $\sim$ 6 grains, consistent with observations in the laboratory and in-situ summarized by \citeA{Rice06}.


\subsection{Initial conditions and slip kinematics}
We next define the initial conditions and the kinematic framework for modeling strain localization and subsequent frictional weakening triggered by a step increase in fault slip velocity.
We illustrate fault gouge evolution near the front of a propagating slow or fast rupture using a simple kinematic analogy. We impose a sudden slip-velocity step applied at a given point on the fault at the time of the passage of the rupture front, $t=0$. 
The slip velocity increases from an ambient value $V_{\mathrm{ini}}$ to a much higher value $V$, with $V\gg V_{\mathrm{ini}}$. Prior to this velocity step, the shear strain rate is assumed to be uniformly distributed within the fault gouge:
\begin{equation}
t\le0:\quad\dot{\gamma}=\dot{\gamma}_{\mathrm{ini}}\equiv \frac{V_{\mathrm{ini}}}{h},\quad f=f_{\mathrm{ini}}\,.
\end{equation}
Post the velocity step  at $t=0$, the point-wise strain rate is generally unknown and subject to the obvious constraint: 
\begin{equation}
t>0:\quad\int_{-h/2}^{h/2}\dot{\gamma}(y,t)\,dy=V
\label{V_int} \,.
\end{equation}
Immediately upon the velocity step, at $t=0^+$, the strain rate increases instantaneously but remains distributed uniformly, since a finite time and strain are required for non-uniform strain to develop within the gouge layer. 
This instantaneous strain-rate increase leads to a uniform increase in friction throughout the gouge layer due to the direct effect:
\begin{equation}
t=0^{+}:\quad\dot{\gamma}=\dot{\gamma}_{0}\equiv \frac{V}{h},\quad f=f_{p}\equiv f_\mathrm{ini}+a\ln \frac{V}{V_\mathrm{ini}}
\label{gamma_0}\,.
\end{equation}
Neglecting poromechanical effects, e.g., dilation \cite{Rice14,brantut2021dilatancy}, within the gouge and surrounding rock, the pore pressure and thus effective normal stress remain initially unchanged:
\begin{equation}
t=0^{+}:\quad p=p_{o},\quad\bar{\sigma}=\bar{\sigma}_{o}\quad (\bar\sigma=\sigma-p_o)\label{p_0}
\end{equation}

Realistically, we would expect that the evolution of strain rate  during localization is affected by perturbations to the otherwise homogeneous initial conditions described above. 
To assess such effects, we modify the initial strain rate by
introducing a small, long-wavelength perturbation as:
\begin{equation}
t=0^{+}:\quad\dot{\gamma}=\dot{\gamma}_{0}+\widetilde{\dot{\gamma}}_{\mathrm{ini}}\cos(2\pi y/h),\quad\tau=\tau_{0} \,,
\label{gamma_pert}
\end{equation}
where the perturbation amplitude is assumed to be small compared to the initial uniform strain rate $\widetilde{\dot{\gamma}}_{\mathrm{ini}}\ll\dot{\gamma}_{0}$.

\subsection{Numerical methodology}\label{sec:numerics}
We numerically solve the coupled frictional, hydraulic, and thermal evolution equations to investigate strain localization and frictional weakening within fault gouge.
Recognizing that non-local effects are prevalent only at the grain-scale and that the localized shear band thickness in a fully non-local analysis is governed by the grain-size, we adopt a simpler, local rheological model that inherently predicts unlimited localization. 
To approximate the finite thickness of the localized shear band in our numerical solutions, we set the minimum grid size equal to the irreducible localized shear-band thickness obtained from the non-local solution for the internal shear-band structure (Eq. \ref{non-loco}), $h^*=2\pi \lambda$.  

Thus, the maximum localized strain-rate in our numerical simulations remains physically bounded, as in nature, reaching a peak value of $\dot{\gamma}_{\mathrm{loc}}\approx V/h^*$ within a single (minimum size) grid spacing of size $h^*$. 
The numerical solution obtained with this ``grid-limited'' local framework is approximately dynamically equivalent to that of the full non-local framework, with our choice of $\lambda=h^*/2\pi$.
This is evidenced by the nearly identical predicted shear-strain-rate distributions outside the localized band in the two frameworks (Fig. \ref{fig:sketch}b), and follows from the matching steady-state gouge strength within the localization band in the two solutions. 

This means that the local strain rate in the grid-limited local framework, $\dot{\gamma}_{\mathrm{loc}}$, matches the non-local strain rate in the non-local framework, $\dot\gamma_\lambda$, within the shear band (Fig. \ref{fig:sketch}c). Likewise, the friction within the band is the same in the two approaches.  As a result, the two modeling frameworks predict nearly identical evolution of the gouge frictional strength and breakdown energy (Suppl. Text S4).

The magnitude of $\dot{\gamma}_{\mathrm{loc}}$ relative to the flash-heating threshold $\dot{\gamma}_{w}=V_{w}/h^*$ determines whether severe flash-heating-induced weakening occurs during localization, as well as the extent of this weakening. 
The evolution of the strain rate $\dot{\gamma}(y,t)$, shear stress $\tau(t)$, and effective normal stress $\bar{\sigma}(y,t)=\sigma_{o}-p(y,t)$ within the fault gouge is governed by Eqs. (\ref{eq}-\ref{tau}), (\ref{df}-\ref{fLV}), (\ref{dp}), and (\ref{V_int}), with initial conditions at $t=0^{+}$ defined by Eqs. (\ref{gamma_0}-\ref{gamma_pert}). 

We solve this problem numerically using a finite difference discretization in space and continuous integration in time. The solution is obtained for the normalized system of governing equations presented in Suppl. Text~S1.   The parameter ranges explored in the numerical simulations are summarized in Suppl. Text~S2. Further details of the numerical method are provided in Suppl. Texts S3 and S4, and the Wolfram Mathematica scripts used in our implementation are made available in the Open Research section. 

\section{Results}\label{sec:results}

In this section, we first describe the general features of the localization process and the associated evolution of gouge strength and fracture energy obtained from the numerical solutions. We follow with an approximate analytical treatment that synthesizes the numerical results and reveals the parametric dependence of the localization process.

\subsection{General features of the localization process from numerical solutions}

Figs.~\ref{fig:tau}-\ref{fig:anatomy} illustrate numerical solutions for the evolution of shear stress (Fig. \ref{fig:tau}), friction, effective normal stress, and shear strain rate inside and outside an emerging localized shear band (Fig. \ref{fig:anatomy}) during ongoing localization in a rate-weakening gouge of thickness $h$ from 1~mm to 1~m. The evolution is qualitatively similar for \textit{fast} ($V=1$ m/s; Figs.  \ref{fig:tau}a and \ref{fig:anatomy}a,c,e) and \textit{slow} ($V=1$ mm/s; Figs.  \ref{fig:tau}b and \ref{fig:anatomy}b,d,f) transients and is characterized by four stages:

(i) \emph{instantaneous} frictional strengthening of the gouge by the direct effect ($\delta=0^+$);

(ii) frictional weakening driven by state evolution toward steady-state sliding with \emph{distributed} shear (well approximated by exponential slip weakening   \cite{AmpueroRubin08,garagash2021} over slip distance $d_c=\gamma_c h$, blue dashed lines in Figs. \ref{fig:tau} and \ref{fig:anatomy}a,b), with  effectively negligible heating and pressurization ($\bar\sigma\approx \bar\sigma_o$, Fig. \ref{fig:anatomy}c,d);

(iii) \emph{abrupt localization} at $\delta\approx\delta_\mathrm{loc}$ (star in Fig. \ref{fig:tau}; see  Suppl. Text S5 for the numerical determination of $\delta_\mathrm{loc}$), characterized by a sharp drop in friction (Fig. \ref{fig:anatomy}a,b) and shear stress (Fig. \ref{fig:tau}), and a similarly abrupt concomitant acceleration/deceleration of shear strain rate inside/outside the emerging localization band (Fig. \ref{fig:anatomy}e,f), respectively;

(iv) \emph{fully localized slip} at high, nearly steady shear strain rate within the shear band (Fig. \ref{fig:anatomy}e,f), producing efficient heating and onset of continuing weakening (Fig. \ref{fig:tau}) by thermal pressurization (Fig. \ref{fig:anatomy}c,d), well approximated by the ``slip-on-a-plane'' TP solution \cite{Rice06} (red dashed lines).

The slip to localization scales approximately with the state-evolution slip distance for distributed shearing, $\delta_\mathrm{loc}\sim \gamma_c h$, for both fast and slow transients and across a wide range of gouge layer thicknesses $h $ (Figs. \ref{fig:tau}-\ref{fig:anatomy}), with fast transients localizing faster at smaller slip. 
The key difference between fast and slow transients is the extent of  localization-driven frictional weakening (Figs. \ref{fig:tau} and \ref{fig:anatomy}a,b). \emph{Fast} transients activate co-localization flash-heating resulting in a dramatic drop of frictional strength to $f_\mathrm{loc,fast}\sim 0.1$, whereas \emph{slow} transients exhibit a more modest co-localization reduction of friction. 

We evaluate the evolution of fracture energy $G$ (breakdown work, \citeA{PalmerRice73,AbercrombieRice05}) from the gouge strength–slip history.
We define $G$ as a function of time $t$ or, alternatively, of accrued slip $\delta=Vt$, or nominal strain $\gamma_{0}=\delta/h$, as
\begin{equation}
G(t)=\int_{0}^{t}[\tau(t')-\tau(t)]V(t')dt' \,.\label{energy}
\end{equation}

Fig. \ref{fig:G} shows $G$ evolution with slip for the same transients as in Fig. \ref{fig:tau}. Fracture energy grows slowly during the pre-localization, distributed-shear stage (ii). This is followed by a finite abrupt co-localization increase to $G_\mathrm{loc}$. We define $G_\mathrm{loc}$ as the cumulative fracture energy of the localization process.  $G_\mathrm{loc}$ scales with the overall friction drop and the localization slip, $G_\mathrm{loc}\sim(f_p-f_\mathrm{loc})\sigma_o\,\delta_\mathrm{loc}$, and thus with  gouge thickness (Figs. \ref{fig:G}, \ref{fig:tau}). 
Post-localization, Fig. \ref{fig:G} shows continuous gradual increase of fracture energy with slip due to weakening by thermal pressurization.

Next, we develop an approximate analytical treatment of these  different stages of the localization process to synthesize the numerical results and expose their parametric dependence.


\subsection{Analytical pre-localization model: distributed shearing}\label{subsec:preloc}

Here, we derive an analytical description of pre-localization stage, characterized by distributed uniform shear. We will show that friction relaxes exponentially over $d_c=\gamma_c h$, while effective stress remains nearly constant, establishing the scaling $\delta_{\mathrm{loc}}\sim\gamma_c h$ and  negligible thermal weakening.
Before the onset of localization, shear-strain and shear-strain-rate (Figs.  \ref{fig:sketch}b, \ref{fig:anatomy}e,f) are approximately uniform across the fault gouge, 
\begin{equation}
\gamma\approx\gamma_{0}=\frac{\delta}{h},\qquad
\dot{\gamma}\approx\dot{\gamma}_{0}=\frac{V}{h},\end{equation}
where index `0' denotes the uniform-shearing solution. 
Following an instantaneous increase upon a velocity step (stage (i) in Fig. \ref{fig:tau}a), the gouge friction relaxes from the peak value $f_{p}$ (Eq. \ref{gamma_0}) toward the residual steady-state value $f_r=f_\mathrm{ss}(\dot{\gamma}_{0})$ (stage (ii) in Fig. \ref{fig:tau}a), uniformly across the layer thickness (Figs.\ref{fig:anatomy}a,b). This relaxation is well described by exponential slip / strain weakening \cite{AmpueroRubin08, garagash2021}, 
\begin{equation}
f\approx f_{0}(\gamma)=f_r+(f_p-f_r)\exp\left( -\frac{\gamma}{\gamma_c}\right)\quad\text{with}\quad f_{p}= f_\mathrm{ini}+a\ln \frac{V}{V_\mathrm{ini}},\quad f_r=f_\mathrm{ss}\left(\frac{V}{h}\right)\,.
\label{f0}
\end{equation}

Early-time evolution of the effective normal stress (pore pressure) is approximately undrained-adiabatic. Neglecting the diffusion transport term in Eq. (\ref{dp}) and integrating taking Eqs. (\ref{tau}) and (\ref{f0}) into account, we can write evolution of effective normal stress in the form resembling the earlier constant-friction Lachenbruch's  \citeyear{Lach80}  expression
\begin{equation}
\bar{\sigma}\approx\bar{\sigma}_{0}(\gamma)= 
\bar\sigma_o\exp\left(-\frac{\gamma}{\gamma_w(\gamma)}\right)\,.
\label{sigma0}
\end{equation}
Here, $\gamma_w$ is the path-averaged weakening strain of the undrained-adiabatic process, defined in terms of the path-averaged friction as
\begin{equation}
    \gamma_w\equiv \frac{\rho c}{\left<f_0\right> \Lambda}\quad\text{with}\quad 
    \left<f_0\right>(\gamma)\equiv \frac{1}{\gamma}\int_0^\gamma f_0(\gamma)d\gamma\,.
\end{equation}

Although $\gamma_w$ depends on $\gamma$, the exponential form (\ref{sigma0}) usefully captures the leading-order decay of $\bar{\sigma}$ with accumulated strain.
The weakening strain has well-defined small and large strain (compared to $\gamma_c$) limits corresponding to a constant $\left<f_0\right>$ approximation by the peak $f_p$ and residual $f_r$ values, respectively. 
The localization strain is a fraction of $\gamma_c$ (see Sec. \ref{Sec:loc} and Fig. \ref{fig:collapse}c), thus, suggesting that it is the peak-friction value of the weakening strain $\gamma_{w,p}=\rho c/(f_p \Lambda)$ that governs the thermal weakening pre-localization. 

For typical gouge hydrothermal parameters ($\rho c\sim3$ MPa/C, $\Lambda\sim$ 0.1 - 1 MPa/C), the thermal weakening strain is $\gamma_w\gg1$, whereas the characteristic strain for evolution of frictional contacts is $\gamma_c\lesssim 1$. 
Thus, the two weakening processes during pre-localization operate on distinct strain scales: friction weakens appreciably over $\mathcal{O}(\gamma_c)$, while thermal weakening remains negligible, as is also evident in our numerical solutions (Figs.~\ref{fig:anatomy}c,d) before localization. 
Consequently, pre-localization dynamics are predominantly driven by the frictional weakening during the distributed gouge slip  (Fig.\ref{fig:anatomy}a,b). i.e.,  the ``low-velocity'' rate-and-state friction response (Eq. \ref{f0}). Conversely, thermal weakening may only become efficient after shear localizes.
Given that flash-heating of friction is predicated on localization, all thermal weakening processes, thermal pressurization and frictional flash-heating are inactive during distributed shearing. 

In summary, prior to localization the evolution of stress with slip $\delta$ or with the nominal strain $\gamma_{0}$ can be approximated by the distributed-shear solution 
\begin{equation}
\delta<\delta_{\mathrm{loc}}:\quad\tau=\tau_{0}(\gamma_{0})=f_{0}(\gamma_{0})\bar{\sigma}_{0}(\gamma_{0}),\quad(\gamma_{0}=\delta/h),\label{tau0}
\end{equation}
where evolution of the friction and effective stress ($f_0$ and $\bar{\sigma}_0$) are given by Eq. (\ref{f0}) and Eq. (\ref{sigma0}), and, because of negligible thermal weakening during distributed slip, $\bar{\sigma}_{0}(\gamma_{0})\approx\bar{\sigma}_{o}$.
The here derived analytical solution which defines the pre-localization ``baseline'' for the next section: a regime governed by ``low-velocity'' rate-and-state friction, where stress evolution follows $f_0(\gamma_0)$ under nearly constant effective stress, setting the initial conditions for the onset of localization. Eq. (\ref{tau0}) will serve as the initial condition for our subsequent localization analysis. 


\subsection{Analytical model of instability of distributed shearing and localization}\label{Sec:loc}
We next develop an analytical model that captures the instability of distributed shearing and predicts the onset of localization. Our analysis first establishes a linear instability criterion, followed by validation of the predicted localization strain and slip against numerical results.

\subsubsection{Linear stability analysis and localization criterion}\label{sec:ls}

We analyze the linear stability of distributed shearing to quantify the growth of small perturbations and to estimate the strain and slip required for shear to localize.  
We consider infinitesimal small perturbations to the baseline solution for uniformly distributed shearing, $\dot{\gamma}_{0}$, $f_{0}(\gamma_{0})$, $\bar{\sigma}_{0}(\gamma_{0})$ with $\gamma_{0}(t)=\dot{\gamma}_{0}t$. 

The such perturbed fields are
\[
\dot{\gamma}=\dot{\gamma}_{0}+\widetilde{\dot{\gamma}}(y,t),\quad f=f_{0}(\gamma_{0}(t))+\widetilde{f}(y,t),\quad\bar{\sigma}=\bar{\sigma}_{0}(\gamma_{0}(t))+\widetilde{\bar{\sigma}}(y,t)
\]
where tildes denote perturbations. Substituting these expressions into the friction evolution equation, Eq. (\ref{df}), and linearizing yields
\begin{equation}
\frac{\partial\widetilde{f}}{\partial t}=\frac{a}{\dot{\gamma}_{0}}\frac{\partial\widetilde{\dot{\gamma}}}{\partial t}-\frac{\dot{\gamma}_{0}}{\gamma_{c}}(\widetilde{f}-\widetilde{f}_\mathrm{ss})-\frac{\widetilde{\dot{\gamma}}}{\gamma_{c}}( f_0(\gamma_{0})-f_r)\,,
\label{f_wigle}
\end{equation}
where $\widetilde{f}_\mathrm{ss}=(a-b)\widetilde{\dot{\gamma}}/\dot{\gamma}_{0}$ is the perturbation of the  steady-state friction assuming the `low velocity' regime as discussed above (Sec. \ref{subsec:preloc}), i.e.,  $f(\gamma_{0})$ is given by Eq. (\ref{f0}).

Because thermal pressurization is negligible during distributed shearing, 
we neglect perturbations to the effective normal stress. Thus, the effective stress remains uniform across the layer. Stress equilibrium then implies that the stress-ratio perturbation is also spatially uniform, as $\partial\widetilde{f}/\partial y=0$. 

Differentiating Eq. (\ref{f_wigle}) with respect to $y$ and using $\gamma_{0}$ as the evolution variable instead of time gives
\begin{equation}
0=a\frac{\partial g}{\partial\gamma_{0}}-\frac{b-a+f_0(\gamma_{0})-f_r}{\gamma_{c}}g \,, \label{f_wigle-1}
\end{equation}
where $g$ is the normalized strain-rate-gradient perturbation, 
\begin{equation}
g\equiv
\frac{1}{\dot{\gamma}_{0}/h} \frac{\partial\widetilde{\dot{\gamma}}}{\partial y} \,,
\label{g-wiggle}
\end{equation}
which is equivalent to the full strain-rate-gradient, given the spatial uniformity of the unperturbed solution.


For small strains, $\gamma_{0}\ll\gamma_{c}$, the solution to Eq.~(\ref{f_wigle-1}) with the initial condition $g(0)=g_\mathrm{ini}$ simplifies to
\begin{equation}
g(\gamma_0)\approx g_{\mathrm{ini}}\exp\left(\frac{\gamma_{0}}{\gamma^{*}}\right)\quad\mathrm{with}\quad \gamma^{*}=\frac{a}{b-a+f_{p}-f_r}\times \gamma_{c}\,.
\label{growth'}
\end{equation}
describing exponential amplification of perturbations
with characteristic strain $\gamma^*$. 

Eq.~(\ref{growth'}) describes the early-stage development of the instability leading to localization. To estimate the nominal strain at localization, $\gamma_{\mathrm{loc}}=\delta_{\mathrm{loc}}/h$, we first extend Eq.~(\ref{growth'}) ad hoc to finite perturbations up to localization, such that $g_\mathrm{loc}\approx g_{\mathrm{ini}}\exp\left(\gamma_\mathrm{loc}/\gamma^{*}\right)$.
We then require the normalized strain-rate-gradient at localization to reach a finite value, i.e., $g_\mathrm{loc}\sim1$. 
Combining with the estimate of the initial strain-rate-gradient from Eqs. (\ref{g-wiggle}) and (\ref{gamma_pert}), gives  $g_\mathrm{ini}\sim\widetilde{\dot{\gamma}}_{\mathrm{ini}}/\dot{\gamma}_{0}$.
We can then write $\exp\left(\gamma_\mathrm{loc}/\gamma^{*}\right)\approx g_\mathrm{loc}/g_{\mathrm{ini}}= \mathcal{O}(1)\times \dot{\gamma}_{0}/\widetilde{\dot{\gamma}}_{\mathrm{ini}}$. 
Solving for $\gamma_\mathrm{loc}$ and substituting the expression for the characteristic strain $\gamma^*$ from Eq. (\ref{growth'}), we obtain
\begin{equation}
\gamma_{\mathrm{loc}}\approx \ln\left(\mathcal{G} \,\frac{\dot{\gamma}_{0}}{\widetilde{\dot{\gamma}}_{\mathrm{ini}}}\right)\,\frac{a}{b-a+f_{p}-f_r}\times \gamma_{c},\quad \delta_\mathrm{loc}=\gamma_\mathrm{loc}h,\quad\left(\mathcal{G}=\mathcal{O}(1)\right).
\label{gamma_loc}
\end{equation}
Equation~(\ref{gamma_loc}) provides an analytical expression for the strain and corresponding slip at which distributed shearing is predicted to localize. The prefactor $\mathcal{G}$, which scales the weak, logarithmic dependence of $\gamma_{\mathrm{loc}}$ on the initial conditions, is chosen below by fitting Eq. ~(\ref{gamma_loc}) to the numerical solutions for localization.

\subsubsection{Validation and scaling of the analytical prediction for localization slip}

We next compare the analytical estimate, Eq.~(\ref{gamma_loc}), with numerical solutions for the localization slip $\delta_\mathrm{loc}$ 
for a wide range of transients, spanning slip velocities from slow-to-fast (100 $\mu$m/s -- 10 m/s), and gouge thicknesses (1~mm -- 1~m) (Fig. \ref{fig:collapse}a). 
The numerically computed localization slip $\delta_\mathrm{loc}$ exhibits an approximately linear scaling with gouge thickness $h$, which is further exposed by 
normalizing the slip as $\delta_\mathrm{loc}/(\gamma_c h)$ (Fig.~\ref{fig:collapse}c), which leads to collapse of three orders of magnitude variation in $\delta_\mathrm{loc}$ to within a factor of two (0.4–1).

Further, normalizing the numerical  $\delta_\mathrm{loc}$ values by the theoretical prediction (Eq.~\ref{gamma_loc}) yields near-complete data collapse (Fig.~\ref{fig:collapse}e) using a fitted prefactor of $\mathcal{G}\approx3$.
 The theoretical and numerical results agree closely for both rate-weakening ($a-b<0$; Fig.~\ref{fig:collapse}a,d,e) and rate-strengthening ($a-b>0$; Suppl. Fig.~S4a,d,e) friction laws. This validation holds for varying prestress $f_\mathrm{ini}$ (captured in $f_p$; Fig.~S5a,d,e), and for a wide range of initial strain-rate perturbations $\widetilde{\dot{\gamma}}_{\mathrm{ini}}$ (Suppl. Fig.~S6a,d,e).

Our analytical expression (Eq. \ref{gamma_loc}) provides a compact physical criterion for the shear localization within the principal slip zone (Fig. \ref{fig:1}a). Distributed deformation is unstable at the onset of slip, leading to exponential growth of strain perturbations.  These culminate in the formation of the principal slip `surface' (Fig. \ref{fig:1}a,b), i.e. the fully-localized shear band. The corresponding values of strain  $\gamma_{\mathrm{loc}}$ and slip $\delta_{\mathrm{loc}}=\gamma_{\mathrm{loc}}h$ at the localization are approximated by Eq. (\ref{gamma_loc}). Our framework links gouge material parameters ($a$, $b$, $\gamma_c$), prestress, gouge thickness and transient slip velocity ($f_p$, $f_r$, $\dot\gamma_0=V/h$), and initial gouge strain-rate heterogeneity ($\widetilde{\dot{\gamma}}_{\mathrm{ini}}$) into a measurable localization threshold. This forms the foundation for our subsequent analysis of post-localization weakening and fracture energy partitioning.


\subsubsection{Stress at localization}
Although the initial small perturbation growth is gradual (Sec.~\ref{sec:ls}), the final stages of localization occur rapidly and, in terms of the stress evolution, can be approximated by an almost instantaneous stress drop at $\delta\approx\delta_{\mathrm{loc}}$ (Fig. \ref{fig:tau}). 
The required near-instantaneous evolution of friction to its fully-localized steady-state value upon localization stems from a very small state-evolution slip distance within the localized band, $d_c=\gamma_c h^*$,  compared to that during pre-localization, $d_c=\gamma_c h$. 
Thus, friction decreases from the low-strain-rate value given by the distributed-shear solution, $f_0(\gamma_{\mathrm{loc}})$ (Eq.~(\ref{f0})), to the steady-state value $f_{\mathrm{loc}}=f_\mathrm{ss}(\dot{\gamma}_{\mathrm{loc}})$ attained at the fully localized strain rate $\dot{\gamma}_{\mathrm{loc}}\approx \!V/h^{*}$.  
During this transition, the effective normal stress remains essentially unchanged from its pre-localization value (Fig. \ref{fig:anatomy}c,d). 

 The state at localization can thus be expressed as
\begin{equation}
\delta=\delta_{\mathrm{loc}}:\quad  \tau_{\mathrm{loc}}\approx f_{\mathrm{loc}}\,\bar{\sigma}_o,\quad f_{\mathrm{loc}}\approx f_\mathrm{ss}(V/h^*)\,.\label{tau_loc}
\end{equation}

For fast transients ($V>V_w$), localization activates flash-heating, resulting in $f_{\mathrm{loc}}\approx f_{w}$ (Fig \ref{fig:tau}a). For slower transients ($V<V_w$), friction $f_{\mathrm{loc}}$ retains a higher, low-strain-rate steady-state value (Fig \ref{fig:tau}b).

\subsection{Analytical model of post-localization thermal weakening: slip-on-a-plane solution}

Following the distributed-shear instability and localization described above, we now analyze the subsequent post-localization weakening driven by thermal pressurization.

After localization, slip proceeds at a nearly constant frictional level $f_{\mathrm{loc}}$, while thermal pressurization progressively reduces the effective normal stress, causing continued weakening. 
This post-localization stage can be described by the classical `slip-on-a-plane' solution, which neglects heat and fluid storativity of the localized shear zone \cite{Rice06,Platt14,ViescaGaragash15}. 
Expressed in terms of post-localization slip $\Delta\delta=\delta-\delta_{\mathrm{loc}}$, the stress evolves as 
\begin{equation}
\delta>\delta_{\mathrm{loc}}:\quad\frac{\tau}{\tau_\mathrm{loc}}\approx \frac{\bar\sigma}{\bar\sigma_o}\approx 
\exp\left(\frac{\Delta\delta}{L^{*}}\right)\,\text{erfc}\sqrt{\frac{\Delta\delta}{L^{*}}},
\label{TP}
\end{equation}
where slip-scale $L^{*}$ is given by
\[
L^{*}=\left(\frac{\rho c}{f_{\mathrm{loc}}\Lambda}\right)^{2}\frac{4\alpha_{\mathrm{hy}}}{V}
\]
As shown by \citeA{Rice06}, this slip-on-a-plane solution predicts continuous, sustained weakening at all slip values characterized by a fractional power-law decay $\tau/\tau_{\mathrm{loc}}\sim\sqrt{L^{*}/(\pi\Delta\delta)}$ at large slip $\Delta\delta\gg L^{*}$.

Fig.  \ref{fig:post}a shows that, when stress (Fig. \ref{fig:tau}) is plotted against slip normalized by $L^{*}$, numerical solutions for a wide range of gouge thicknesses $h$ (1~mm -- 1~m) collapse onto a single curve.  
The theoretical prediction (Eq. \ref{TP}), parametrized using our theoretical solutions for localization slip $\delta_\mathrm{loc}$ (Eq. \ref{gamma_loc}), and stress $\tau_\mathrm{loc}= f_\mathrm{loc}\,\bar\sigma_o$ (Eq. \ref{tau_loc}), shown by red dashed lines in Fig. \ref{fig:post},  provides an excellent approximation for the collapsed post-localization numerical solutions 
for both fast and slow transients. 

The transition to localization marks a sharp, rate-dependent drop of fault strength controlled by the reduced state-evolution slip distance $\sim h^{*}$ and the activation of flash heating at high slip rates. Once localized, slip proceeds under efficient thermal pressurization, driving continued weakening characterized by the TP slip scale $L^{*}$. Together, Eqs.~(\ref{tau_loc})–(\ref{TP}) provide a quantitative framework linking the frictional transition at $\delta_{\mathrm{loc}}$ to the ensuing continuous slip-dependent weakening and energy dissipation during post-localization slip.

\subsection{Analytical model of fracture energy decomposition}

Lastly, we develop analytical approximations for the fracture energy (Eq. \ref{energy}) at different stages of slip, pre-, during and post-localization. We validate these results against numerical solutions. 
Our analysis follows from Eq.~(\ref{energy}) using the approximate strength evolution derived above, assuming negligible thermal pressurization prior to localization.

\subsubsection{Fracture energy of distributed and localized shearing}

Pre-localization, the fracture energy associated with distributed shearing follows from Eq.~(\ref{energy}) with the stress–strain relation (Eq. \ref{tau0}), as
\[
\delta<\delta_{\mathrm{loc}}:\quad G\approx\frac{(f_{p}-f_r)\bar{\sigma}_{o}}{2\gamma_{c}}\frac{\delta^{2}}{h}\,.
\]
This quadratic dependence reflects the approximately linear weakening of friction with strain during uniform shearing (Eq. \ref{f0}).

At localization, the total fracture energy  (shown by the opaque-blue area in Fig. \ref{fig:tau}) includes both 
the distributed pre-localization component evaluated at $\delta=\delta_{\mathrm{loc}}$, and an additional increment
due to the nearly instantaneous strength drop from distributed
to  localized shear. 
The total energy of the localization process is therefore
\begin{equation}
    G_{\mathrm{loc}}\approx\Delta f_{\mathrm{loc}}\bar{\sigma}_{o}\delta_{\mathrm{loc}}\,,
    \label{G_loc}
\end{equation}
where 
\[
\Delta f_{\mathrm{loc}}\approx \left(f_{p}-\frac{1}{2} (f_{p}-f_r)\frac{\gamma_{\mathrm{loc}}}{\gamma_{c}}\right)-f_{\mathrm{loc}}
\]
is an equivalent friction drop of the localization process. 
For fast slip transients ($V>V_w$), $f_{\mathrm{loc}}\approx f_{w}$, and therefore, to the leading order, $\Delta f_{\mathrm{loc}}\approx f_{p}-f_{w}$ corresponds to the drop from peak friction at the onset of sliding to flash-heated friction during localized shearing (Fig. \ref{fig:tau}a). 
The localization fracture energy $G_{\mathrm{loc}}$ therefore scales approximately linearly with localization slip $\delta_{\mathrm{loc}}\sim h$, implying an approximate proportionality between gouge thickness and the  fracture energy dissipated during localization.

\subsubsection{Post-localization fracture energy and decomposition}

Post-localization weakening, governed by thermal pressurization (TP, Eq. \ref{TP}), leads to continued growth of the total  
fracture energy with slip (Fig. \ref{fig:G}).  From Eq.~(\ref{energy}) and the slip-on-a-plane solution, we can write 
\begin{equation}
\delta>\delta_{\mathrm{loc}}:\quad G\approx G_{\mathrm{loc}}+\Delta G(\Delta\delta)\label{energy_decomp}
\end{equation}
where the incremental fracture energy of TP weakening is given by
\[
\Delta G(\Delta\delta)=\int_{0}^{\delta_{\mathrm{loc}}}[\tau_{\mathrm{loc}}-\tau(\Delta\delta)]d\delta'+\int_{0}^{\Delta\delta}[\tau(\Delta\delta')-\tau(\Delta\delta)]d(\Delta\delta'),\quad\Delta\delta=\delta-\delta_{\mathrm{loc}} ,
\]
where $\tau(\Delta\delta)$ is given by the 'slip-on-a-plane' TP solution (Eq. \ref{TP}).

Evaluating this integral yields 
\begin{equation}
\Delta G(\Delta\delta)=\left[1-\frac{\tau(\Delta\delta)}{\tau_{\mathrm{loc}}}\right]\tau_{\mathrm{loc}}\delta_{\mathrm{loc}}+\left[2\sqrt{\frac{1}{\pi}\frac{\Delta\delta}{L^{*}}}+\left(1-\frac{\Delta\delta}{L^{*}}\right)\frac{\tau(\Delta\delta)}{\tau_{\mathrm{loc}}}-1\right]\tau_{\mathrm{loc}}L^{*}\,.
    \label{DG}
\end{equation}
The first term in this expression for $\Delta G$ 
saturates at large slip ($\tau_{\mathrm{loc}}\delta_{\mathrm{loc}}$), while the second term grows continuously, producing the sustained fracture energy increase associated with thermal pressurization. 
At large slip, $\Delta G\!\sim\!\tau_{\mathrm{loc}}L^{*}\sqrt{\Delta\delta/(\pi L^{*})}$, indicating a fractional power-law dependence on slip \cite{Rice06}.
Equation~(\ref{energy_decomp}) thus decomposes the total fracture energy into (i)~the gouge-thickness-dependent localization energy, $G_{\mathrm{loc}}\!\propto\!h$, and (ii)~the slip-dependent post-localization component $\Delta G(\Delta\delta)$ with the slip-scale $L^{*}$.

\subsubsection{Validation and scaling of fracture energy}
The analytical results are validated against numerical simulations in Fig.~\ref{fig:G} and Fig.~\ref{fig:collapse} (b,d,f).  

Figure ~\ref{fig:G} compare numerical solutions for the fracture energy evolution with slip in fast, (a), and slow, (b), transients with the theoretical predictions for the localization fracture energy  $G_{\mathrm{loc}}$ and the post-localization fracture energy $G_\mathrm{loc}+\Delta G(\delta-\delta_\mathrm{loc})$, shown by blue and red dashed lines for the case of gouge thickness $h=1$ m, respectively. These predictions are based on Eqs. (\ref{G_loc}) and (\ref{DG}), respectively, evaluated using analytical solutions for $\gamma_{\mathrm{loc}}$ and $\delta_{\mathrm{loc}}$  (Eq.~\ref{gamma_loc}), and $\tau_{\mathrm{loc}}$ (Eq. \ref{tau_loc}).

Figures~\ref{fig:collapse}b and~\ref{fig:collapse}d show that the localization fracture energy  $G_\mathrm{loc}$ evaluated from numerical solutions scales approximately linearly with gouge thickness $h$ across both fast and slow transients. When the localization fracture energy is normalized by the theoretical prediction based on Eq. (\ref{G_loc}) the data collapse over three orders of magnitude in $h$ (Fig.~\ref{fig:collapse}f).  
The same scaling behavior is observed for rate-strengthening friction (Suppl. Fig.~S4b,d,f), varying prestress levels (Fig.~S5b,d,f), and different perturbation magnitudes (Suppl. Fig.~S6b,d,f), confirming the robustness of the analytical model.

In summary, our derived expressions establish that the total fracture energy can be decomposed into a fault-size-dependent minimum component, $G_{\mathrm{loc}} \propto h$, and a slip-dependent component, $\Delta G(\Delta\delta)$, arising from post-localization thermal weakening. This framework quantitatively links the mechanics of shear localization and the efficiency of subsequent thermal weakening to the fault energy budget, providing a physical basis for the observed scaling of fracture energy with both the fault size and the coseismic slip \cite{AbercrombieRice05,ViescaGaragash15,gabriel2024_FractureEnergy}.

\section{Discussion} 

\subsection{From distributed shear in the principal slip zone to a localized principal slip surface}
Our results suggest a robust, three-phase evolution (Fig.~\ref{fig:tau}a) of co-seismic shear strength when slip in a fluid-saturated gouge is driven by a rapid velocity increase (passing front of a slow or fast rupture).
Slip begins delocalized within the principal slip zone of thickness $h$ and weakens slowly under classical rate–state friction, potentially aided by weak, undrained thermal pressurization. Once the frictional state has evolved by a strain $\sim \gamma_c$, slip abruptly localizes at the grain/asperity scale $h^*$ triggering flash heating if the slip is fast enough; thereafter, fully localized slip continues to weaken by efficient thermal pressurization on an effectively mathematical plane. This cascade holds for both steady-state rate-weakening and  rate-strengthening gouges; in the latter, the transient state evolution still drives instability and localization (see Suppl. Figs. S1-S4). 

\subsection{Fracture-energy partitioning tied to gouge thickness and co-event slip}
The above evolution provides a natural \emph{fracture-energy} partition:
\begin{equation}
G(\delta) \;=\; G_\mathrm{loc}(h) \;+\; \Delta G(\Delta\delta),
\label{eq:Gdecomp}
\end{equation}
where the \emph{localization} component $G_\mathrm{loc}$ accumulates up to the slip at localization, $\delta_{\mathrm{loc}}$, and the \emph{post-localization} component $\Delta G$ accumulates thereafter due to TP weakening during ``slip-on-a-plane'' $\Delta\delta=\delta-\delta_\mathrm{loc}$. 

A key outcome is that
\begin{equation}
\delta_{\mathrm{loc}} \;\sim\; \gamma_c\,h,
\qquad\Longrightarrow\qquad
G_\mathrm{loc} \;\propto\; h,
\end{equation}
such that the minimum fracture energy needed to form a fully localized principal slip surface scales approximately linearly with the principal slip zone thickness $h$ (and thus with fault maturity/size; Fig.~\ref{fig:observations}). 

The strong flash-heating-induced friction drop at localization contributes to $G_\mathrm{loc}$ when the peak localized strain rate, $\dot\gamma_{\mathrm{loc}}\!\approx\!V/h^*$, exceeds the flash-heating threshold $\dot\gamma_w\approx 10^4$. 
It thereby effectively demarcates the fracture energy of localization for fast (flash-heated) and slow (non-flash-heated) slip transients, with $G_\mathrm{loc,fast}/G_\mathrm{loc,slow}\approx 2.5$ for rate-weakening ``low-velocity'' friction (Fig.~\ref{fig:collapse}b, Suppl. Figs. S5 and S7) and $\approx 7$ for rate-strengthening ``low-velocity'' friction (Suppl. Fig. S4b). 
The thickness of the fully-localized shear band, $h^*$, and therefore the median gouge grain size, clearly affect the flash-heating activation ($\dot\gamma_\mathrm{loc}\ge \dot\gamma_w$).
Coarse gouges (larger $h^*$) require higher slip velocity $V$ to activate flash-heating. 
Our default  value $h^*=10\,\mu\text{m}$, based on natural ultracataclastic gouges with $\sim 1\,\mu\text{m}$ median grain size, implies $V_w\approx 0.1$ m/s and thus flash-heating activation over a wide range of seismic slip transients with $V>0.1$ m/s. 
Wider shear bands, $h^*\sim100\,\mu\text{m}$, as observed in laboratory experiments \cite{Proctor14,YaoMaPlatt15} imply  $V_w\sim 1$ m/s  and therefore flash-heating activation only for strong seismic transients with $V\gtrsim 1$m/s (Suppl. Fig. S7). 

After localization, the weakening path collapses onto the TP slip-on-a-plane solution and becomes largely independent of $h$.  Thermal pressurization supplies sustained weakening with a characteristic slip scale $L^*$ that is independent of~$h$ and set by hydrothermal properties and $V$, implying that $\Delta G(\delta)$ increases with slip and progressively dominates for large~$\delta$ (Fig.~\ref{fig:G}).

Together, these results reconcile small-event fracture energy, dominated by $G_\mathrm{loc}(h)$, with large-event behavior where $\Delta G(\Delta\delta)$ renders $G$ increasingly fault-invariant with growing slip \cite{gabriel2024_FractureEnergy}.

\subsection{Rectifying the range of the state-evolution strain $\gamma_c$}

The state evolution strain is the key gouge property controlling the localization strain, slip and fracture energy. It is therefore important to verify that our assumed range, $\gamma_c\!\sim\!0.1$--$1$, is consistent with observed localization strains in frictional and compressional 
failure tests, when interpreted through our localization model. 
Biaxial undrained compression of dense sands yields $\gamma_{\mathrm{loc}}\!\sim\!0.1$ \cite{HanVardoulakis91,MokniDesrues1999}, whereas triaxial compression of intact rocks indicates $\gamma_{\mathrm{loc}}\!\sim\!0.01$ \cite{Rummel78}. 
Using the approximate relation $\gamma_{\mathrm{loc}}\!\sim\!(0.1$--$0.5)\,\gamma_c$ emerging from our analysis, these measurements are consistent with $\gamma_c\!\sim\!0.1$--$1$. 
This range ensures that $\delta_{\mathrm{loc}}\!\sim\!(0.1\text{--}0.5)\gamma_c h$ remains small compared with typical coseismic slip on mature faults, yet large enough to carry a measurable $G_\mathrm{loc}$ that scales with~$h$. 
It also ensures that thermal pressurization, which becomes efficient only at the much larger strain scale $\gamma_w \gg1$, has negligible influence on the localization process, in contrast to previous studies that neglect frictional state evolution in the gouge.

\subsection{Relation to previous localization models}
Our results extend and revise earlier analyses of gouge localization under thermal-pressurization weakening. 
Prior kinematic and dynamic treatments of distributed shear with a \emph{rate-only} (velocity-strengthening) frictional rheology found localization to be driven by thermal pressurization at a strain $\gamma_\mathrm{loc}$ comparable to the undrained-adiabatic characteristic strain $\gamma_w\!=\!\rho c/(f\Lambda)\!\gg\!1$ \cite{Rice14,Platt14,platt2015TDlocalization,barrasbrantut2025localization, WangHeimisson2026PoroelasticLocalization}.
By incorporating the full rate--\emph{and}--state response, we show that the transient state evolution is the primary driver of localization, yielding a much smaller localization strain, $\gamma_{\mathrm{loc}}\sim\gamma_c\lesssim1$, negligible TP effects \textit{prior} to localization, and drastically smaller $\delta_{\mathrm{loc}}$ and $G_\mathrm{loc}$ than inferred previously with rate-only rheologies. 
Our conclusions regarding the driver of localization, namely the evolution of frictional contact state in the gouge leading to apparent mechanical weakening with slip or strain, are more consistent with studies that invoke similar gouge-weakening mechanisms, including pressure-sensitive, non-associative, hardening-softening plasticity \cite{MuhlhausVardoulakis1987LocalizationCosserat, Vardoulakis1991gradient,  rattez2018localization_theory,rattez2018localization_num,stathas2023localization} and shear-transformation-zone 
inelastic theory \cite{daub2008_STZ_localization}.


In our framework, TP becomes decisive \emph{after} localization, governing $\Delta G(\Delta\delta)$, whereas FH produces the abrupt friction drop at localization and increases $G_\mathrm{loc}$ without affecting $\delta_\mathrm{loc}$) in fast transients. This framework also clarifies why localization occurs irrespective of whether the low-velocity steady-state friction branch is rate-weakening or rate-strengthening (Suppl. Figs. S1-S4), since the non-steady state evolution behind the rupture front provides the apparent weakening required to localize strain. In contrast, steady-state rate strengthening behavior eventually promotes delocalization in the tail of slow transients that fail to activate flash heating (Suppl. Fig. S2f).




\subsection{Heterogeneity, repeated ruptures, and (de-)localization within the principal slip zone (PSZ)}
Natural PSZs are heterogeneous (e.g., the SAFOD core), and initial slip within a low-velocity \emph{strengthening} segment may be partitioned across sub-bands within a delocalized PSZ. 



Inter-event \emph{healing}, the homogenization of gouge strength and structure, will control whether a subsequent transient must re-localize or instead inherits a previously formed principal slip surface. Where slip transients occur frequently (e.g., as triggered or episodic slow slip events), 
healing may be incomplete, favoring persistent localization across events. 
Conversely, substantial interseismic aging promotes initially delocalized slip, requiring renewed localization during each fast event \cite{beeler2022DcAgeDependent}. Resolving these competing tendencies is essential for assessing the generality and persistence of co-event localization in mature faults.



\subsection{Implications for earthquake scaling and observables}

Our approximately linear scaling $G_\mathrm{loc}\!\propto\!h$, exemplified by $G_\mathrm{loc}\approx 10\text{ [MPa]}\times\gamma_c\,h$ in Fig. \ref{fig:collapse}b and Suppl. Fig. S5b, implies a fault-size dependence of the \emph{small-slip} fracture energy because $h$ increases with both cumulative fault displacement $D$ and fault size $R$. 
Using $h\!\approx\!10^{-2}D$ for unsaturated growth (Fig.~\ref{fig:1}c) and $D\!\approx\!10^{-2}R$ \cite{dawers1993FaultScaling, Scholz10}, the predicted $G_\mathrm{loc}\approx 1000 \text{ [Pa]}\times \gamma_c\,R$ scales linearly with fault dimension and helps explain seismological inferences of an approximately linear lower-bound $G$-vs.-size trend for small earthquakes and small-slip sub-events,  $G_c\approx 400\text{ [Pa]}\times R$  (Fig.~\ref{fig:observations}, and Gabriel et al. \citeyear{gabriel2024_FractureEnergy}). 
Full agreement between the localization theory prediction $G_\mathrm{loc}$ and the observed lower-bound $G_c$ is obtained for a realistic value of the state-evolution-strain, $\gamma_c\approx 0.4$. 
At large slip, the fault-size-invariant contribution $\Delta G(\Delta\delta)$ increasingly dominates, consistent with observations that $G$ becomes less sensitive to fault size and more sensitive to coseismic slip in large crustal and subduction events \cite{AbercrombieRice05, ViescaGaragash15}. 
Microstructural evidence for extreme coseismic shear localization within narrow PSSs embedded in wider PSZs \cite{Chester98,chester2005fracture,Rice06} supports the mechanistic plausibility of this scaling.





\subsection{Future work}
The localization solutions developed here for an advancing rupture front are kinematic, based on slip-rate-step solutions.  Extending this work to fully elasto-dynamic models (e.g., \cite{barrasbrantut2025localization, WangHeimisson2026PoroelasticLocalization, daub2010_STZ_Rupture}) could clarify the subsequent co-event evolution of localized slip, the potential for co-event de-localization, and ultimately the inter-event evolution of fault gouge and fast and slow fault slip cycles. 
In extending this work to rupture elastodynamics, our simplified numerical treatment of the finite thickness of the localized band through  grid-limiting will need to be replaced by an explicit non-local granular rheology. 
Although we show that this simplified approach is approximately equivalent to a non-local approach based on \cite{pranger2022RS} under the simplifying assumptions of a kinematically imposed slip velocity step acting on a homogeneous principal slip zone gouge, natural PSZ heterogeneity,  possible disruption along-fault PSZ continuity in large faults \cite{Faulkner10}, possible shear-band (principal slip surface) migration \cite{platt2015TDlocalization, stathas2023localization} and/or multiple shear banding \cite{garagash2006LocalizationPatterns} within the PSZ gouge, and potential delocalization \cite{Faulkner10, savage2024localization}  all call for fully non-local approaches on an unbiased mesh.

Dynamically incorporating such non-local effects, together with dilatancy, poroelastic coupling, and hydrothermal diffusion along the fault \cite{WangHeimisson2026PoroelasticLocalization}, as well as realistic evolution of hydrothermal properties \cite{BrantutPlatt2017FH,badt2023TP_path, WangHeimisson2026PoroelasticLocalization}, may refine estimates of $(\delta_{\mathrm{loc}},G_c)$ and of the onset and efficiency of flash heating and thermal pressurization. 
Finally, coupling this localization physics to fully dynamic rupture or seismic-cycle models that incorporate realistic fault-core and principal-slip-zone structure and heterogeneity \cite{Chester04, Faulkner10, MitchellFaulkner09, savage2011faultzoneobservation,gabriel2021DiffuseFault} could quantify how often natural ruptures re-use pre-existing principal slip surfaces versus (re-)localize within the principal slip zone, and how this choice controls macroscopic earthquake characteristics such as rupture speed, high-frequency radiation, and event-to-event variability.



\section{Conclusions}
%

We show that localization in gouge is a \emph{transiently state-driven} process that (i) sets a minimum fracture-energy budget,  $G_\mathrm{loc}(h)$,  that depends on principal-slip-zone thickness,  and (ii) triggers rapid flash heating followed by efficient thermal pressurization, which together provide a co-transient, \emph{slip-dependent} contribution $\Delta G(\Delta\delta)$. This decomposition unifies field, laboratory, and seismological observations across small and large events, and it provides a concrete mechanistic pathway from initially distributed shear to a co-seismic principal slip surface.

\begin{figure*}
\includegraphics[width=1.1\textwidth]{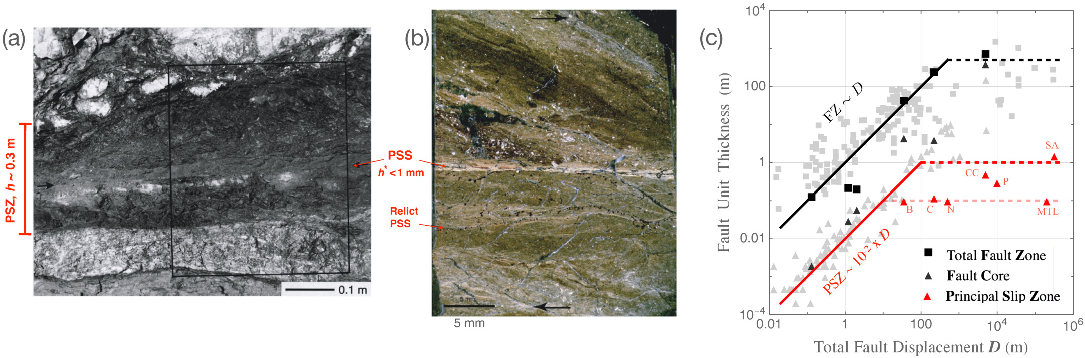}
\caption{ 
(a) Principal Slip Zone (PSZ, ~ $h\sim 0.3$ m) of a large fault (the Punchbowl fault), corresponding to a partially low-cohesion ultracataclastic gouge unit \cite{Chester98}. A much thinner Principal Slip Surface (PSS, ~ $h^*<1$ mm) has been inferred to accommodate most coseismic slip. 
(b) A thin section sample in cross-polarized light shows the PSS and further extreme shear localization, with an apparent thickness of 100~$\mu$m \cite{chester2005fracture, Rice06}. The PSS may result from coseismic localization of slip initially distributed across a wider PSZ. The mechanical breakdown work of this coseismic localization process, which contributes to the total fracture energy of an earthquake, is hypothesized to scale with the PSZ thickness. 
(c) Scaling of Principal Slip Zone (PSZ, triangles \cite{MitchellFaulkner08,Mizoguchi08,zoback2011scientific,Chester98,WibberleyShimamoto03}) and Fault Zone (FZ, squares \cite{savage2011faultzoneobservation,MitchellFaulkner08,marrett1990kinematic,robertson1987fault}) thickness with total fault displacement $D$, a measure of fault maturity. For large faults, the PSZ is defined as the ultracataclastic gouge layer (red traingles), whereas for smaller faults and fractures it is defined as the fault core (FC, grey and black triangles). 
See text and Table~1 for details. Labels denote SA - San Andreas at SAFOD; P - Punchbowl; MTL - Median Tectonic Line; CC - Caleta Coloso; N - Nojima; C - Cristales; B - Blanca faults. Solid lines show linear fits, whereas dashed lines indicate approximate saturation of the linear scaling at fault displacements greater than $\sim 10 -100$ m for the PSZ and $\sim 500$ m for the FZ.
Total fault displacement scales the principal slip zone thickness, which in turn scales coseismic localization slip and fracture energy.
}
\label{fig:1}
\end{figure*}

\begin{figure*}
\includegraphics[width=\textwidth]{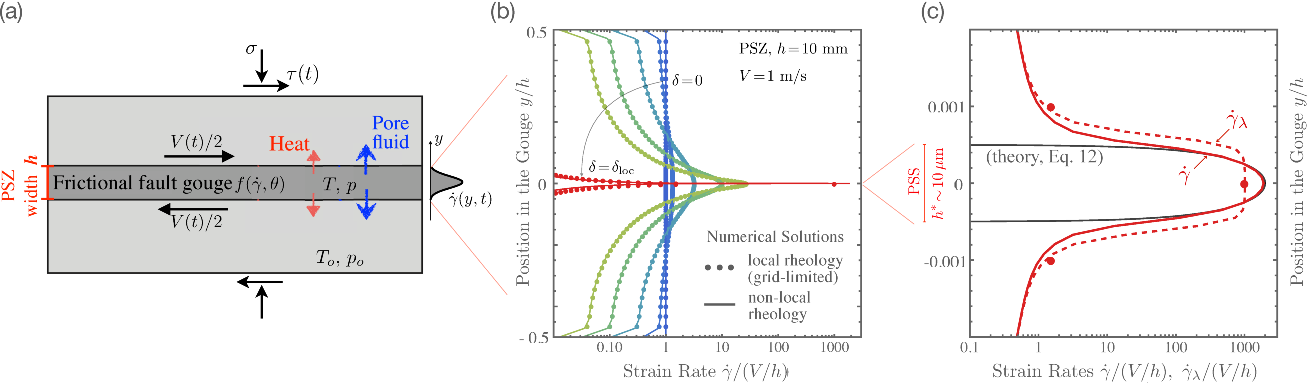}
\caption{ 
Strain localization within a gouge layer, or principal slip zone (PSZ), is driven by non-steady frictional relaxation behind the rupture front, modeled as a slip velocity step in time. Localization activates thermal weakening processes that produce both a near-instantaneous (flash-heating) and more extended (thermal pressurization) fault weakening with slip (see also Fig. 3).
(a) Schematic of PSZ with thickness $h$, sheared at slip rate $V$. (b) Example of the strain-rate evolution across a PSZ of thickness $h=10$ mm as slip evolves following a co-seismic velocity step  $V\gg V_\mathrm{ini}$. The initially nearly uniformly distributed slip localizes onto a thin co-seismic principal slip ``surface'' (PSS) of thickness $h^*\sim$10 microns over a slip distance $\delta_\mathrm{loc} \sim \gamma_c \, h$. The frictional state-evolution strain $\gamma_c$  is a gouge property inferred to fall between 0.01 to 1 (see text). (c) Zoomed view of panel (b), showing the structure of the PSS once formed.
Numerical solutions in (b-c) are shown for local (filled circles) and non-local (lines) frictional rheologies formulated in terms of the local strain rate $\dot\gamma$ and a mollified non-local strain rate $\dot\gamma_\lambda=(1+\lambda^2\,\partial^2/\partial y^2)\,\dot\gamma$, respectively, where the micro-structural (grain) scale $\lambda$ is related to the final PSS thickness, $\lambda=h^*/2\pi$ (see text). The two solutions are indistinguishable at the scale of the PSZ, (b), whereas non-local effects become apparent within the PSS, (c).  
}
\label{fig:sketch}
\end{figure*}

\begin{figure*}
\includegraphics[width=\textwidth]{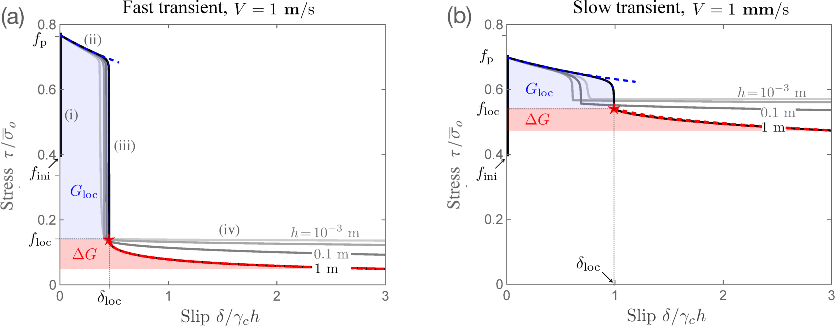}
\caption{Evolution of transient \textit{shear strength}, scaled by the initial effective normal stress, with \textit{slip} scaled by $\gamma_c h$, following a step change of slip velocity to $V=1$ m/s in (a), and $V=1$ mm/s in (b), within gouge layers of thickness $h=$ 1, 10, 100 mm (opaque gray lines), and $h=$ 1 m (black line) undergoing localization.
The localization slip  $\delta_\mathrm{loc}$ and stress $\tau_\mathrm{loc}\approx f_\mathrm{loc}\bar\sigma_o$ are marked by a star for the case $h=1$ m.  The blue dashed line (all $h$) and red dashed line ($h=1$ m) show approximate analytical solutions for, respectively, the frictional weakening with slip during distributed shear and the TP-on-a-plane weakening with slip \cite{Rice06} during fully localized sliding. The former pre-localization weakening is governed by relaxation of the frictional state with nominal strain $\delta/h$, Eq. (\ref{f0}), which leads to an approximate collapse of the stress--strain response independent of gouge thickness. The latter post-localization weakening is governed by TP-driven reduction of effective normal stress with slip, Eq. (\ref{TP}), rather than with strain, such that the  post-localization stress vs. nominal strain evolution for different gouge thicknesses approximately collapses onto an invariant stress vs. excess- slip response (Fig. \ref{fig:post}) characterized by a single hydrothermal slip-scale $L^*$.
Fig. \ref{fig:anatomy} shows the corresponding evolution of friction, effective normal stress, and strain rate at the center of the gouge layer, where deformation localizes, and contrasts it with the evolution outside the localization band at the boundary of the gouge layer.
}
\label{fig:tau}
\end{figure*}

\begin{figure}
\includegraphics[width=\textwidth]{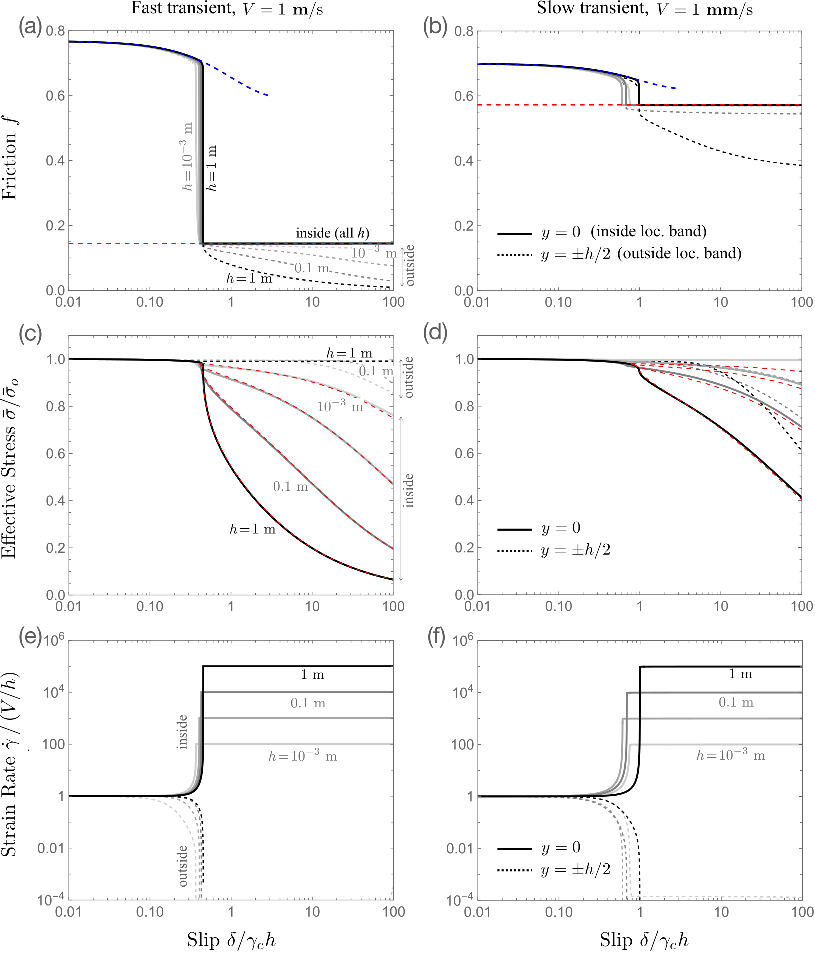}
\caption{Evolution of friction, effective normal stress scaled by its initial value, and strain rate scaled by the nominal strain rate $V/h$, at the gouge center line ($y=0$, where strain localizes) and at the gouge layer boundaries ($y=\pm h/2$), with slip scaled by $\gamma_c h$, for the fast (a,c,e) and slow (b,d,f) transients. Blue and red dashed lines show the analytical pre- and post- localization solutions, respectively.}
\label{fig:anatomy}
\end{figure}

\begin{figure*}
\includegraphics[width=\textwidth]{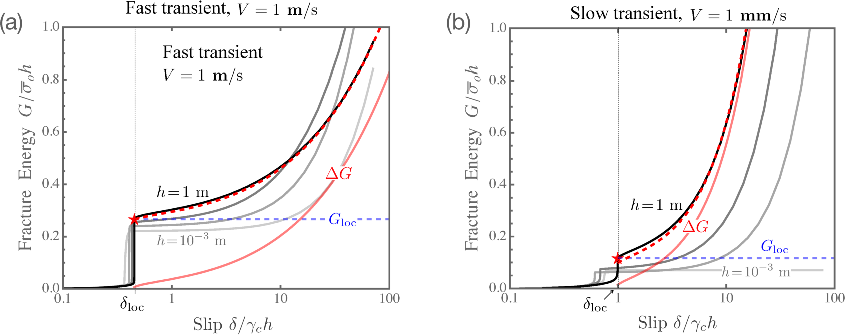}
\caption{Evolution of the \textit{fracture energy}, scaled by $\bar\sigma_o h$, with \textit{slip} scaled by $\gamma_c h$, for the fast (a) and slow (b) transients shown in Fig. \ref{fig:tau}. The results suggest an approximate post-localization decomposition, $G = G_\mathrm{loc} + \Delta G$ (red dashed line for $h=1$ m case), into the fracture energy of localization, $ G_\mathrm{loc} \sim (\tau_p-\tau_w) \delta_{loc}$ (blue dashed line), which is proportional to gouge thickness through $\delta_{loc}\sim \gamma_c h$, and the fracture energy associated with continued TP weakening during post-localization slip,  $ \Delta G(\delta-\delta_{loc})$, well-approximated by the TP-on-a-plane solution \cite{Rice06}, opaque red line, which is independent of gouge thickness and continues to increase with slip.}
\label{fig:G}
\end{figure*}

\begin{figure*}
\includegraphics[width=\textwidth]{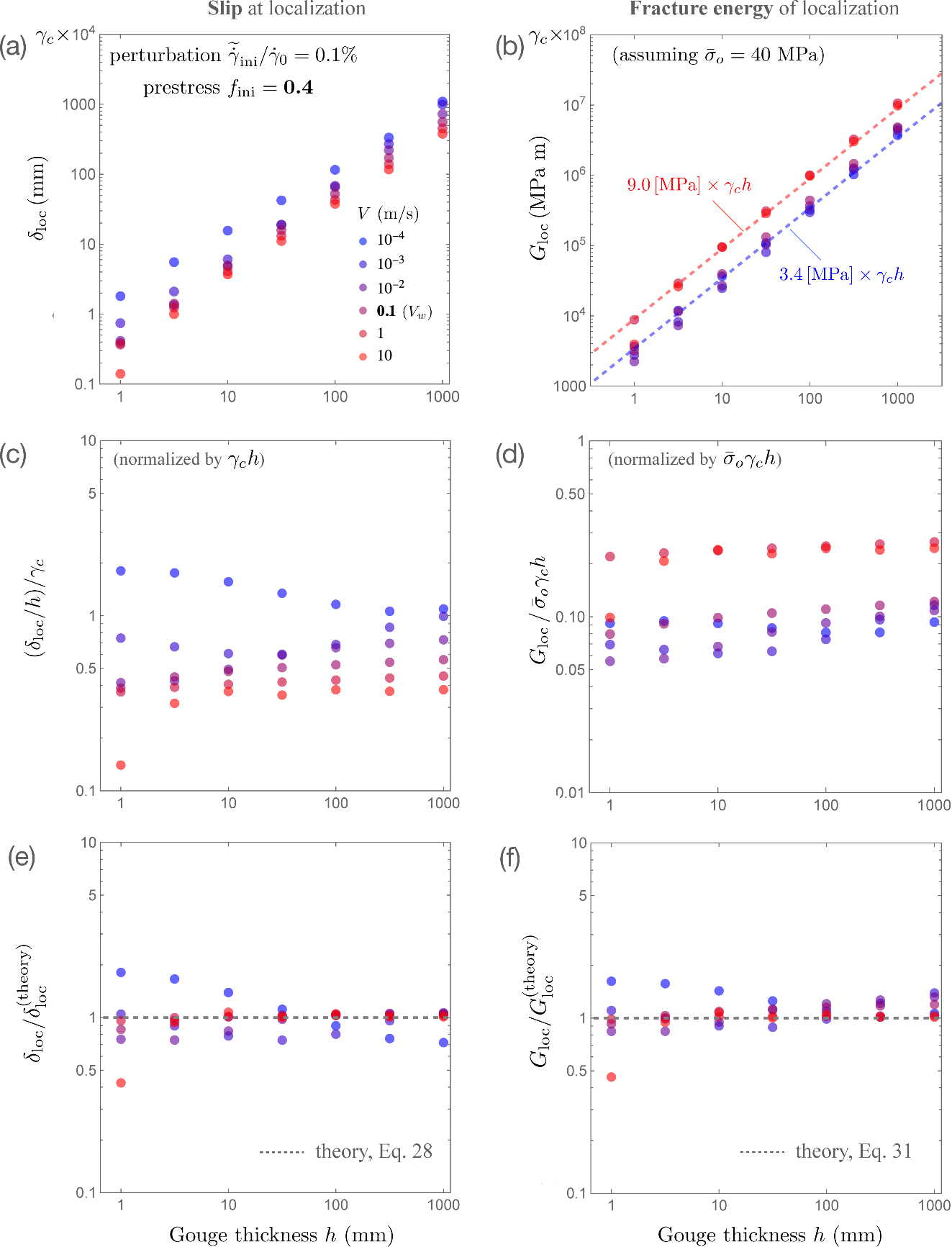}
\caption{Slip to the shear localization $\delta_\mathrm{loc}$ (a,c,e) vs. the gouge layer thickness $h$ varying over several orders of magnitude, representing transition from immature, small faults with small cores to large plate bounding faults (Fig. 1a), in response to a step change of the fault slip velocity to a value $V$ in the range spanning from slow $10^{-4}$ to fast $10$ m/s (color-coded). The fracture energy (breakdown work) of the localization process $G_\mathrm{loc}$ is shown in (b,d,f). The gouge rate-and-state dependent friction is steady-state rate weakening (RW) at low velocities. (Solutions for low-velocity rate-strengthening (RS) gouge are in Sup. Fig. S4). The initial fault prestress ratio is $f_\mathrm{ini}=0.4$. (Solutions for smaller and larger prestress are in Sup. Fig. S5). (a,b) show dimensional slip and energy scaled by the value of the gouge state evolution strain $\gamma_c$  varying approximately linear with gouge thickness for a given transient strength (value of $V$). The two linear fits for the fracture energy corresponding to fast ($V\ge1$ m/s) and slow ($V\le 0.1$m/s) transients corresponding to whether or not slip is fast enough to activate flash heating weakening upon localization are included in (b). (c,d) show that the localization slip and fracture energy are approximately (velocity-dependent) constants when normalized by the gouge thickness. (e,f) show the localization slip and fracture energy \textit{collapse} when normalized by the approximate theoretical solution given by Eqs. (\ref{gamma_loc}) and (\ref{G_loc}), respectively.  
}
\label{fig:collapse}
\end{figure*}

\begin{figure}
\includegraphics[width=\textwidth]{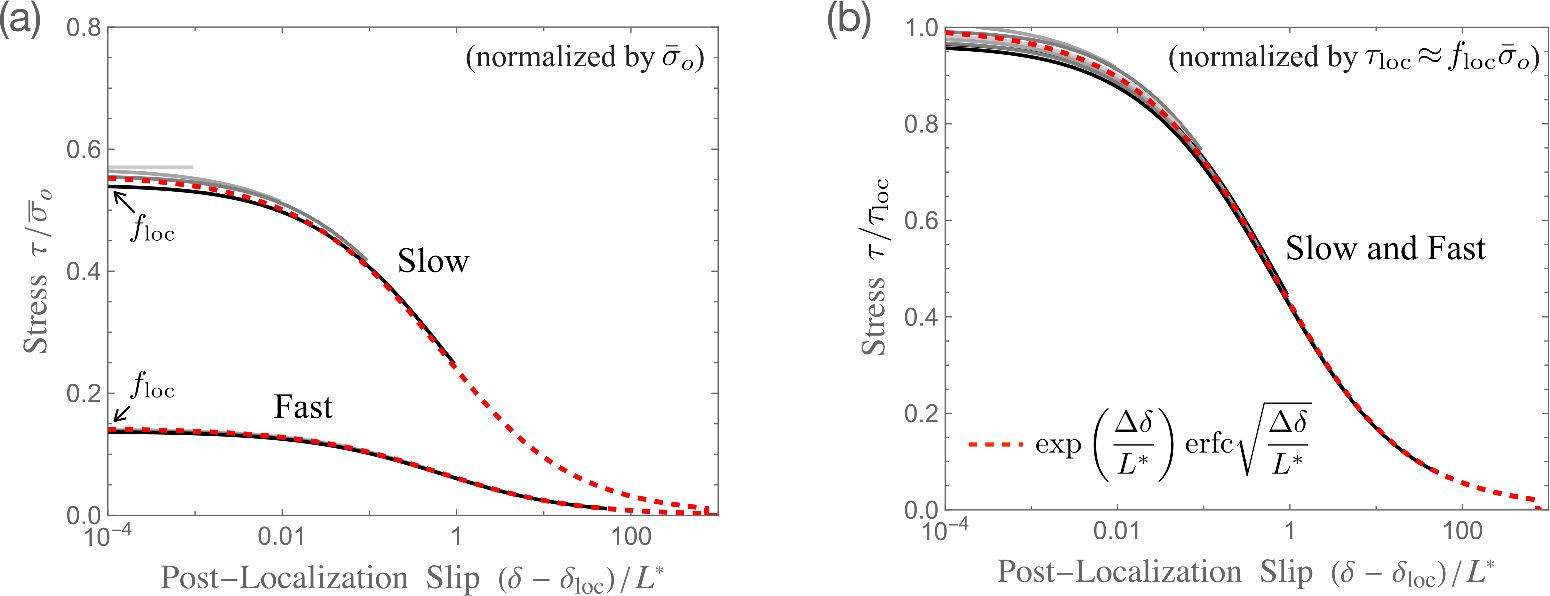}
\caption{
\textit{Post-localization} evolution of transient \textit{shear strength} as a function of post-localization slip for the fast and slow transients over a wide range of gouge thicknesses (Fig. \ref{fig:tau}).  Panel (a) shows the collapse of shear strength onto TP slip-on-a-plane solution with constant friction (Eq. (\ref{TP})) adopted from Rice \citeyear{Rice06}, when expressed in terms of the excess slip $\delta-\delta_\mathrm{loc}$ scaled by hydrothermal, slip-rate-dependent scale $L^*$. In contrast to the pre-localization strength evolution, which is  dominated by the initial gouge thickness $h$ (Fig. \ref{fig:tau}), the post-localization strength is approximately independent of $h$ and is parametrized by the two end-member values of the friction at localization, $f_\mathrm{loc}$, corresponding to fast, flash-heated and slow, non-flash-heated transients (stress-axis intercepts in panel a).
Panel (b) shows that the post-localization strength for both fast and slow transients collapses onto a single master curve when normalized by  its value at localization, $\tau_\mathrm{loc}\approx f_\mathrm{loc}\bar\sigma_o$ (Eq. \ref{tau_loc}).
}
\label{fig:post}
\end{figure}

\begin{figure*}
\includegraphics[width=0.75\textwidth]{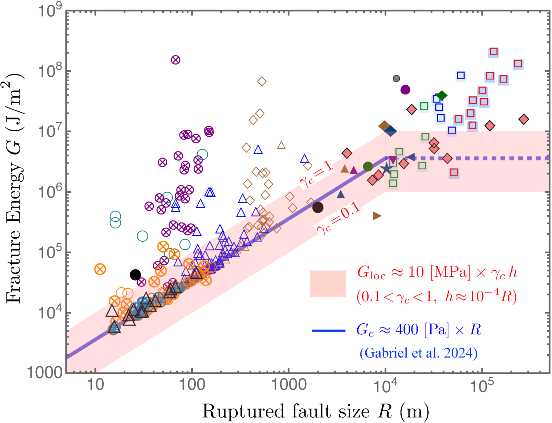}
\caption{The observed approximately linear fault-size dependence of fracture energy for ``small-slip'' events can be explained by the fracture energy of localization on faults whose ``maturity'' scales with PSZ gouge-layer thickness (Fig. 1c). 
Symbols show seismological inferences of fracture energy of small crustal (open), large crustal (solid), and large subduction (filled squares) earthquakes, as well as 3D earthquake simulations of past well-recorded earthquakes (pink-filled diamonds and black circles) compiled by Gabriel et al. \citeyear{gabriel2024_FractureEnergy} and references therein. The blue solid line shows their theoretical prediction for the lower-bound fracture energy, $G_c(R)\approx 400\text{ [Pa]}\times R$, based on a crack-like, flash-heated rupture. 
The pink-shaded band shows the predicted dependence of localization fracture energy on PSZ gouge thickness, $G_\mathrm{loc}(h)\approx 10\text{ [MPa]}\times \gamma_c\,h$ (Fig. \ref{fig:collapse}b, Suppl. Fig. S5b), for frictional state evolution strain $\gamma_c$ ranging from $0.1$ to $1$, and assuming that the PSZ gouge thickness depends linearly on fault size $h(R)\approx 10^{-4} R$. The PSZ thickness scales linearly with cumulative fault displacement, $h\approx 10^{-2} D$ (Fig. 1c), which itself scales linearly with fault size, $D\approx 10^{-2} R$ \cite{dawers1993FaultScaling,Scholz10}). This scaling saturates at sufficiently large cumulative fault slip ($D>100$ m), or equivalently at large fault size ($R>10$ km). 
The predicted scaling of localization fracture energy matches that of the lower-bound fracture energy inferred by Gabriel et al. \citeyear{gabriel2024_FractureEnergy}, i.e.,  $G_\mathrm{loc}(h(R))\approx G_c(R)$, for a realistic value of the state-evolution strain $\gamma_c\approx 0.4$.
}
\label{fig:observations}
\end{figure*}

In detail, we find that: 
\begin{enumerate}
\item  In faults governed by rate-and-state friction, slip transients initially distribute shear across the fault gouge and subsequently localize it to the asperity or grain scale. This transition is a fundamental mechanism governing earthquake rupture.
\item The slip distance to localization, $\delta_\mathrm{loc}$, scales approximately linearly with gouge thickness $h$, and increases weakly with transient velocity $V$ and with decreasing amplitude of the initial strain-rate perturbation within the gouge. The associated fracture energy, $G_\mathrm{loc}$, scales similarly to $\delta_\mathrm{loc}$, but with a marked dichotomy in its dependence on $V$, depending on whether the transient is strong enough to activate flash-heating.
\item Localization leads to continued fault weakening with post-localization slip $\Delta\delta$, and thus to a 
continuously increasing fracture-energy contribution $\Delta G(\Delta\delta)$.
 Rapid slip enhances thermal weakening through (i) flash heating at frictional grain contacts within the shear band, which occurs nearly instantaneously upon localization and contributes to the localization fracture energy $G_\mathrm{loc}$; and (ii) delayed thermal pressurization, together with extended flash-heating weakening after localization, modulated by heat and fluid diffusion. The latter drives continued weakening with slip and the slip-dependent fracture-energy increment $\Delta G(\Delta\delta)$.
\item Coupled with the scaling of fault gouge thickness $h$ with fault size $R$ (Fig. \ref{fig:1}), this localization theory provides a physical mechanism underlying the decomposition of coseismic fracture energy into $G = G_c(R) + \Delta G(\delta)$, proposed by Gabriel et al. \citeyear{gabriel2024_FractureEnergy} based on seismological observations, in which the fault-size-dependent component can be understood as $G_c(R)=G_\mathrm{loc}(h(R))$.
\item In our model, localization occurs on both rate-weakening and rate-strengthening faults. The underlying driver is the apparent weakening with slip associated with evolution of the frictional state variable in response to a fault slip-rate perturbation, from the peak friction associated with the direct effect to a residual, steady-state value. Neglecting state-evolution in some of the previous localization models leads to strong overestimation of both the localization slip and the associated fracture-energy contribution.
\end{enumerate}


\newpage

\begin{table}[h]
    \centering
    \caption{
    Thickness of the PSZ (ultracataclasite gouge layer) is approximately saturated $\sim 0.1-1$ m for large faults with total displacement $\gtrsim 10 - 100$ m (Fig. \ref{fig:1}).}
    \label{tab:1}
    \begin{tabular}{l c c lp{9cm}}
        \hline
        Fault & Total Slip (km) & PSZ (m)&  Reference
(Slip)&Reference
(PSZ)\\
        \hline
        San Andreas 
& 315& 1.5 &  Atwater \citeyear{Atwater1998SAF}&Zoback et al.
\citeyear{zoback2011scientific}\\
 (at SAFOD)& & & &(Fig 4, geophisical logs and caliper data)\\
 MTL& $>200$& 0.1 & Wibberley \& Shimamoto
\citeyear{WibberleyShimamoto03}&Wibberley \& Shimamoto
\citeyear{WibberleyShimamoto03}\\
        Punchbowl         & $>10$& 0.3 &  Chester \& Chester \citeyear{Chester98}&Chester \& Chester \citeyear{Chester98}\\
        Nojima& 0.5& 0.1 &  Murata et al. 
\citeyear{murata2001Nojima}&Mizoguchi et al.
\citeyear{Mizoguchi08}\\
        Caleta Coloso     & 5.     & 0.5 &  Mitchell \& Faulkner
\citeyear{MitchellFaulkner09}&Mitchell \& Faulkner
\citeyear{MitchellFaulkner09}\\
        Cristales         & 0.22   & 0.12 &  Mitchell \& Faulkner
\citeyear{MitchellFaulkner09}&Mitchell \& Faulkner
\citeyear{MitchellFaulkner09}\\
        Blanca            & 0.035  & 0.1 &  Mitchell \& Faulkner
\citeyear{MitchellFaulkner09}&Mitchell \& Faulkner
\citeyear{MitchellFaulkner09}\\
        \hline
    \end{tabular}
\end{table}

\newpage
\vspace*{\fill}

\vspace*{\fill}\clearpage

\newpage
\vspace*{\fill}

\vspace*{\fill}\clearpage

\newpage
\vspace*{\fill}

\vspace*{\fill}\clearpage

\newpage
\vspace*{\fill}

\vspace*{\fill}\clearpage

\newpage
\vspace*{\fill}

\newpage
\vspace*{\fill}

\vspace*{\fill}\clearpage

\section*{Open Research}
The Wolfram Mathematica scripts containing the implementation of the numerical method, examples of numerical solutions and post-processing can be downloaded from \url{https://github.com/dgaragash/Localization-and-fracture-energy}.

\section*{CRediT
}
Conceptualization: DIG, AAG \\
Data curation: DIG\\
Formal analysis: DIG\\
Funding acquisition: DIG, AAG\\
Investigation: DIG, AAG\\
Methodology: DIG\\
Resources: DIG, AAG\\ 
Software: DIG\\
Validation: DIG, AAG\\
Visualization: DIG\\
Writing – original draft: DIG, AAG

\acknowledgments

We would like to thank Harsha S. Bhat, Ake Fagereng and Casper Pranger for stimulating discussions.
DIG acknowledges support by the Natural Science and Engineering Research Council of Canada (NSERC) (Discovery grant no. 05743). 
AAG acknowledges support by the National Science Foundation (NSF) (grant nos. EAR-2121568, EAR-2225286, OAC-2311208, OAC-2139536, RISE-2531036), by Horizon Europe (ChEESE-2P, grant number 101093038, DT-GEO, grant number 101058129, and Geo-INQUIRE, grant number 101058518), the Statewide California Earthquake Center (SCEC, grants no. 25341, 25313), and the National Aeronautics and Space Administration (NASA, grant no. 80NSSC20K0495).

\newpage

\bibliography{bibrefs_garagash_Aug21, Alice-Refs}

\end{document}